\documentclass[useAMS,usenatbib]{mn2e}
\usepackage{graphicx}
\usepackage{amssymb}
\usepackage{subfigure}
\usepackage{booktabs}
\usepackage{tabularx}
\usepackage{url}
\usepackage{threeparttable}
\bibpunct[; ]{(}{)}{,}{a}{}{,}

\def\xmm{{\it XMM-Newton}}
\def\SNR{\mbox{{MCSNR~J0508--6902}}}

\newcommand{\SII}{[S\,{\sc ii}]}
\newcommand{\NII}{[N\,{\sc ii}]}
\newcommand{\OIII}{[O\,{\sc iii}]}
\newcommand{\Halpha}{H${\alpha}$}
\newcommand{\vapec}{{\it vapec}}

\newcommand{\vsedov}{{\it vsedov}}

\title[Study of a new LMC SNR, \SNR]
  {Multi-frequency study of a new Fe-rich supernova remnant in the Large Magellanic Cloud, \SNR}
\author[L. M. Bozzetto et al.]
  {L. M.~Bozzetto,$^1$ P. J.~Kavanagh,$^2$ P.~Maggi,$^3$ M. D.~Filipovi\'c,$^1$ M.~Stupar,$^{4,5}$  
  \newauthor Q. A.~Parker,$^{4,5,6}$ W. A.~Reid,$^{4,5}$ M.~Sasaki,$^2$  F.~Haberl,$^3$ D.~Uro\v{s}evi\'c,$^{7,8}$  
  \newauthor J.~Dickel,$^9$ R.~Sturm,$^3$ R.~Williams,$^{10}$ M.~Ehle,$^{11}$ R. Gruendl,$^{12}$ Y.-H. Chu,$^{12}$ 
  \newauthor S. Points,$^{13}$ \& E. J.~Crawford$^1$
\\
  $^1$University of Western Sydney, Locked Bag 1797, Penrith South DC, NSW 1797, Australia\\
  $^2$Institut f$\rm\ddot{u}$r Astronomie und Astrophysik T$\rm\ddot{u}$bingen, Universit$\rm\ddot{a}$t T$\rm\ddot{u}$bingen, Sand 1, 72076 T$\rm\ddot{u}$bingen, Germany \\
  $^3$Max-Planck-Institut f\"{u}r extraterrestrische Physik, Giessenbachstra\ss e, D-85748 Garching, Germany\\
  $^4$Department of Physics, Macquarie University, Sydney, NSW 2109, Australia\\
 Ê$^5$Australian Astronomical Observatory, PO Box 296, Epping, NSW 1710, Australia\\
  $^6$Research Centre for Astronomy, Astrophysics and Astrophotonics, Macquarie University, Sydney, NSW 2109 Australia\\
  $^7$Department of Astronomy, Faculty of Mathematics, University of Belgrade, Studentski trg 16, 11000 Belgrade, Serbia\\
  $^8$Isaac Newton Institute of Chile, Yugoslavia Branch\\
  $^9$Physics and Astronomy Department, University of New Mexico, MSC 07-4220, Albuquerque, NM 87131, USA\\
  $^{10}$Coca-Cola Space Science Center, 701 Front Avenue, Columbus, GA 31901\\
  $^{11}$XMM-Newton Science Operations Centre, ESAC, ESA, PO Box 78, 28691 Villanueva de la Ca\~nada, Madrid, Spain\\
  $^{12}$Department of Astronomy, University of Illinois, 1002 West Green Street, Urbana, IL 61801, USA\\
  $^{13}$Cerro Tololo Inter-American Observatory, Casilla 603, La Serena, Chile
  }
\date{Released 2011 Xxxxx XX}

\pagerange{\pageref{firstpage}--\pageref{lastpage}} \pubyear{2002}

\def\LaTeX{L\kern-.36em\raise.3ex\hbox{a}\kern-.15em
    T\kern-.1667em\lower.7ex\hbox{E}\kern-.125emX}

\begin{document}

\label{firstpage}

\maketitle

\begin{abstract}
We present a detailed radio, X-ray and optical study of a newly discovered Large Magellanic Cloud (LMC) supernova remnant (SNR) which we denote \SNR. Observations from the Australian Telescope Compact Array (ATCA) and the \xmm\ X-ray observatory are complemented by deep H$\alpha$ images and Anglo Australian Telescope AAOmega spectroscopic data to study the SNR shell and its shock-ionisation. Archival data at other wavelengths are also examined. The remnant follows a filled-in shell type morphology in the radio--continuum and has a size of $\sim$74~pc $\times$ 57~pc at the LMC distance. The X-ray emission exhibits a faint soft shell morphology with Fe-rich gas in its interior -- indicative of a Type~Ia origin. The remnant appears to be mostly dissipated at higher radio-continuum frequencies leaving only the south-eastern limb fully detectable while in the optical it is the western side of the SNR shell that is clearly detected. The best-fit temperature to the shell X-ray emission ($kT = 0.41^{+0.05}_{-0.06}$~keV) is consistent with other large LMC SNRs. We determined an O/Fe ratio of $<21$ and an Fe mass of 0.5--1.8~$M_{\sun}$ in the interior of the remnant, both of which are consistent with the Type~Ia scenario. We find an equipartition magnetic field for the remnant of $\sim$28~$\mu$G, a value typical of older SNRs and consistent with other analyses which also infer an older remnant. 

\end{abstract}

\section{Introduction}

The study of supernova remnants is an imperative part in our understanding of the universe, as they have a profound effect on their surrounding environment. SNRs distribute heavy elements throughout the galaxy, with thermonuclear (Type~Ia) explosions contributing a significant amount of iron into the surrounding medium, whilst core-collapse supernovae (SNe) inject large amounts of Oxygen and other $\alpha$ elements \citep{1986A&A...154..279M}. Type~Ia SNe are thought to arise when a C-O white dwarf (in a binary system) approaches the Chandrasekhar limit. This occurs either through accretion from the binary companion, or by a merger with a white dwarf companion, and is currently actively debated (see e.g. \citealt{2008MNRAS.384..267M} and references therein). Core-collapse SNe are believed to be the result of a progenitor star (with a main sequence mass above 8 solar masses) being drained of its nuclear fuel and therefore not being able to hold up against its own gravity through the release of nuclear energy.

The Large Magellanic Cloud (LMC) is an irregular dwarf galaxy located in close proximity (50 kpc, \citealt{2008MNRAS.390.1762D}) to our own Galaxy, allowing modern instruments to reach high sensitivity levels with sufficient angular resolution, from radio to X-ray wavelengths. At this distance, and because of its moderate inclination angle of 35$^\circ$ \citep{2001AJ....122.1807V}, we are able to assume objects which lay within it are all at a similar distance. This allows us to analyse the surface brightness and spatial extent of an object accurately, which is often difficult in our own Galaxy due to uncertainties in distance, foreground absorption, and background confusion. Furthermore, the LMC is considered an almost ideal environment for the study of a wide range of celestial objects due to the presence of active star-forming regions and its location outside of the Galactic plane, where absorption by gas and dust is reasonably low.

Several multi-wavelength SNR selection criteria exist in the literature (e.g. \citealt{1985ApJ...292...29F}) as well as the powerful, newly evaluated emission-line diagnostic criteria reported by \citet{2010PASA...27..129F} and updated in \citet{2013MNRAS.431..279S}. These are based on the well-known \citet[hereafter SMB]{1977A&A....60..147S} SMB \Halpha/\NII\ versus \Halpha/\SII\ diagnostic diagram and the \citet[hereafter BPT]{1981PASP...93....5B} BPT \OIII5007/H$\beta$ versus \SII/\Halpha~diagnostic diagrams. In this paper we adopt the criteria used by the Magellanic Cloud Supernova Remnant (MCSNR) Database\footnote{See \url{http://www.mcsnr.org/about.aspx}} which are effectively compatible with these publications. In this classification system, a SNR must satisfy at least two of the following three observational criteria: significant \Halpha, \SII, and/or \OIII\ optical line emission with an observed \SII/\Halpha\ flux ratio of $>$ 0.4 \citep{1973ApJ...180..725M}; extended non-thermal radio emission indicative of synchrotron radiation \citep{1998A&AS..130..421F} and extended thermal X-ray emission. 

The ROSAT position sensitive proportional counter (PSPC) catalogue of X-ray sources in the LMC \citep{1999A&AS..139..277H} contains the object [HP99] 791. While this source was confirmed as extended, its large off-axis angle prevented a reliable classification at that time. In this paper, we present new radio-continuum and X-ray observations of [HP99] 791, from the Australia Telescope Compact Array (ATCA) and \xmm, respectively. We also present for the first time the clear identification of an unambiguous complete optical shell for the remnant from a deep arcsecond-resolution \Halpha\ image, a multi-exposure stack of the central 25 square degrees of the LMC \citep{2006MNRAS.365..401R}. The shell can also be seen at lower resolution in the Magellanic Clouds Emission Line Survey (MCELS) data. Finally, we present new optical spectra from several components on the enhanced western side of the optical shell taken by the AAOmega spectrograph on the AAT \citep{2004SPIE.5492..389S,2006SPIE.6269E..14S} which confirm a strong signature of shock excitation as expected for an SNR. This new discovery clearly satisfies all three criteria for identification as a bona-fide SNR and so we assign to the source the identifier \SNR\ following the MCSNR Database nomenclature.

The observations used in this project are described in Section~2, followed by the data analysis in Section~3. The results from these observations are explained in Section~4 and then discussed in Section~5.

\section{Observations}

\subsection{Radio}

The ATCA is a radio interferometer located in northern New South Wales, Australia. Its new 2~GHz bandwidth and improved sensitivity has made it possible to obtain observations of a previously poorly studied group of SNRs, the older and more evolved SNRs that were not detected so far. We observed \SNR\ on the 15$^\mathrm{th}$ and 16$^\mathrm{th}$ of November 2011 with the ATCA, using the new compact array broadband backend (CABB) receiver in the array configuration EW367, at wavelengths of 3 and 6~cm ($\nu$=9000 and 5500~MHz). Baselines formed with the $6^\mathrm{th}$ ATCA antenna were omitted, as the other five antennas were arranged in a compact configuration. The observations were carried out in the so called `snap-shot' mode, totalling $\sim$50 minutes of integration over a 14 hour period. The source PKS~B1934-638 was used for primary (flux) calibration and the source PKS~B0530-727 was used for secondary (phase) calibration. The phase calibrator was observed for a period of 2 minutes every half hour during the observations. 

In addition to our own observations, we made use of a 36~cm (843~MHz) Molonglo Synthesis Telescope (MOST) mosaic image (as described in \citealt{1984AuJPh..37..321M}), a 20~cm (1400~MHz) mosaic image \citep{2007MNRAS.382..543H} as well as 6~cm (4800~MHz) and 3~cm (8640~MHz) mosaic images from project C918 \citep{2005AJ....129..790D}.

\subsection{X-ray}

\citet{1999A&AS..139..277H} detected this source in their compilation of ROSAT PSPC observations, recording an X-ray position of RA(J2000)=05$^h$08$^m$36.7$^s$, DEC(J2000)=--69\degr02\arcmin54\arcsec, and an extent of 43.6\arcsec. It is listed as [HP99]~791. However, due to the large off-axis angle of the source, a classification was not possible. 

We observed \SNR\ with \xmm\ on two occasions: in May 2010 (Obs. ID 0651880201, PI M. Sasaki), and, more recently, in September 2012 (Obs. ID 0690752001, PI F. Haberl) as part of the LMC survey. The European Photon imaging Camera (EPIC) onboard \textit{XMM-Newton} \citep{Jansen2001} consists of a pn CCD \citep{Struder2001} and two MOS CCD imaging spectrometers \citep{Turner2001}. The observational data were reduced using the standard reduction tasks of SAS\footnote{Science Analysis Software, see \url{http://xmm.esac.esa.int/sas/}} version 12.0.1, filtering for periods of high particle background. Unfortunately, Obs. ID 0651880201 was blighted by very high background, resulting in only $\sim9$ ks of the exposure being useful. However, in the case of Obs. ID 0690752001, there was virtually no background contamination with almost all data ($\sim25$~ks) available for further analysis.

\subsection{Optical: Imaging and Spectroscopy}

\begin{figure}
\centering\includegraphics[trim=0 100 0 120,clip,scale=.4]{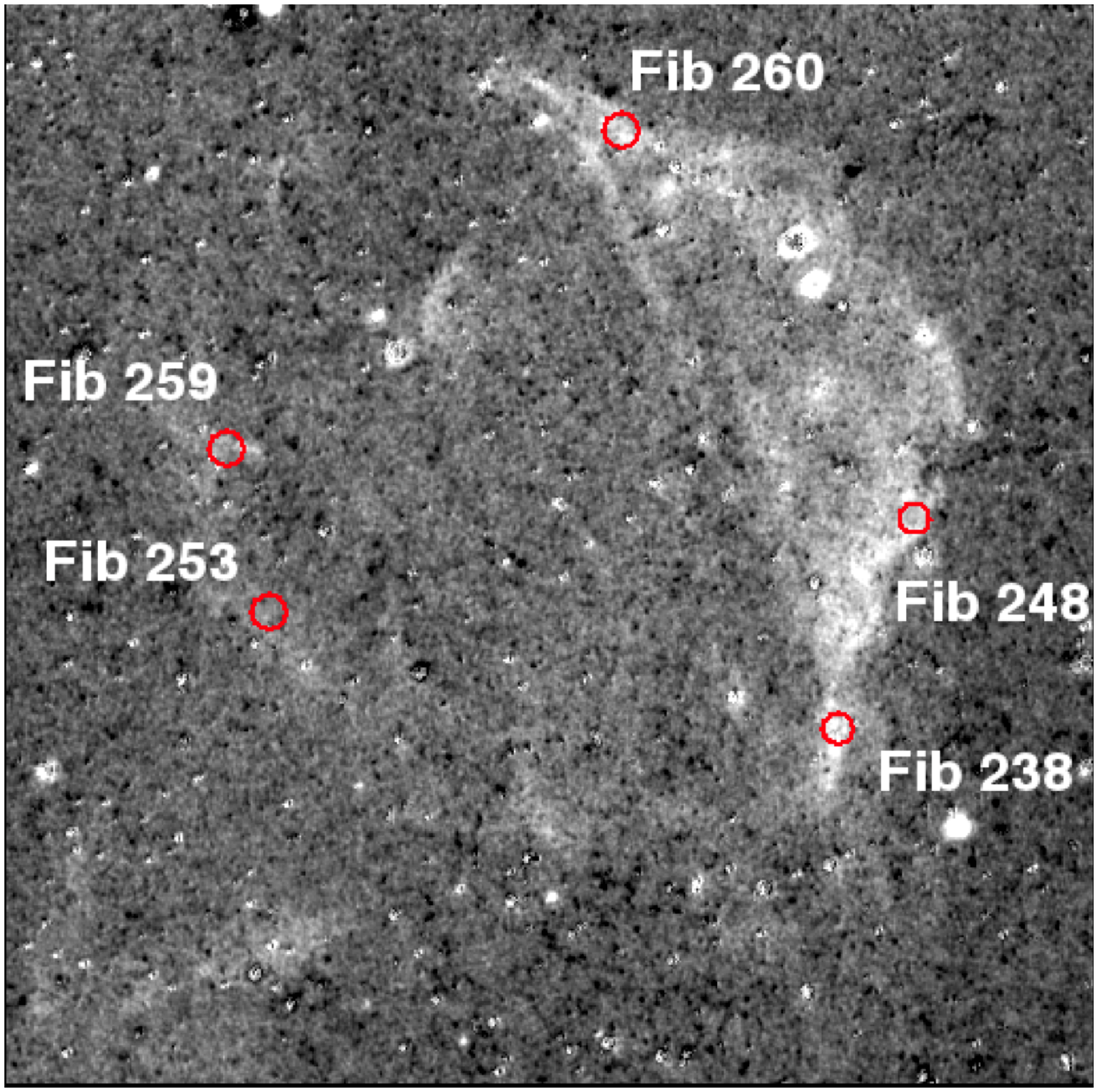}
\caption{\Halpha\ quotient image of the new SNR \SNR\ taken from the data of \citet{2006MNRAS.365..401R} revealing the complete optical oval shell for the first time but with enhanced emission on the western side. The positions of the 2dF fibres for the optical spectroscopy are indicated. The image is $\sim6\arcmin \times 6$\arcmin\ on a side and north-east is to the top left.
 \label{HAQ}}
\end{figure}

We have used deep, arcsecond-resolution, narrow-band optical \Halpha\ images created from a median stack of a dozen 2~hour exposures on the central 25~square degrees of the LMC as described in detail by \citet{2006MNRAS.365..401R}. These data have been used as the basis for unprecedented new discoveries of emission line sources in general and Planetary Nebulae (PN) in particular across the central LMC (e.g. \citealt{2006MNRAS.365..401R,2006MNRAS.373..521R,2012MNRAS.425..355R}). The survey filter, detector and other parameters are described by \citet{2005MNRAS.362..689P} but in essence the LMC multiple exposures were taken with fine-grained Kodak Tech-Pan films on the UK Schmidt Telescope using a monolithic interference filter of exceptional quality (e.g. \citealt{1998PASA...15...33P}). These exposures were scanned with the SuperCOSMOS measuring machine \citep{2001MNRAS.326.1295H}, (included with an in-built accurate WCS) transformed into electronic form and then median stacked to reach depths of R$_{equiv}$$\sim$22 for \Halpha\ (4.5$\times10^{-17}$ erg~cm$^{-2}$ s$^{-1} {\mbox \AA}^{-1}$). We extracted \Halpha\ and matching SR (short `broad-band' red) fits images over a $6\arcmin\times6\arcmin$ area centered on the SNR with 10 $\mu$m (0.67") pixels from the median-stacked images to create a quotient image (\Halpha\ divided by the SR) as shown in Fig.~\ref{HAQ} reproduced to clip to the inner 98\% of intensity values. North-east is to the top left. A complete oval shell can be seen in \Halpha\ but with the western side being enhanced. The centroid of this oval shell and hence the best estimate for the optical centre of the remnant is RA(J2000)=05$^h$08$^m$33.7$^s$ and DEC(J2000)=--69\degr02\arcmin33\arcsec. The major axis is 4.5\arcmin\ in extent and the minor axis only 4\arcmin. The positions of the 2dF spectral samples obtained are also indicated on the figure (see below for further details). 

\begin{figure}
\centering\includegraphics[angle=-90,trim=0 10 0 0,scale=.41]{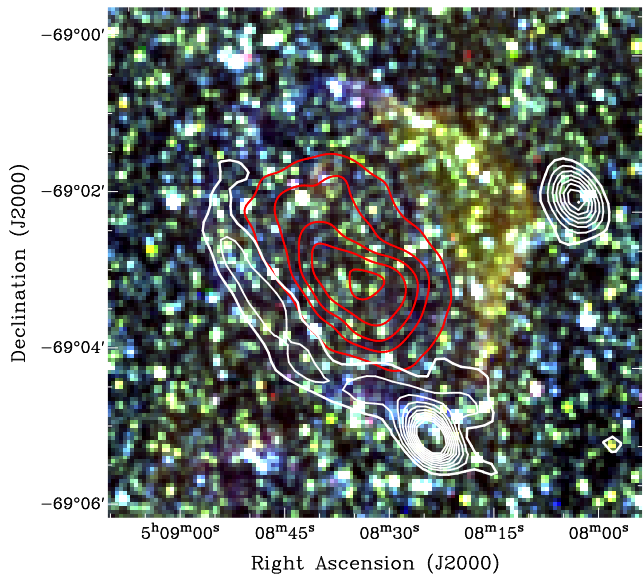}
\caption{MCELS composite optical image \textrm{(RGB =H$\alpha$,[S\textsc{ii}],[O\textsc{iii}])} of \SNR\ overlaid with \xmm\ contours (0.7--1.1~keV; red) of 7, 14, 21, 28 \& 35$\times$10$^{-6}$ cts/s and ATCA (5500~MHz; white) contours from 3$\sigma$ to 30$\sigma$ in steps of 3$\sigma$.
\label{xmmmcels}}
\end{figure}

We also checked for detection of the SNR in the Magellanic Cloud Emission Line Survey (MCELS) that was carried out with the 0.6~m University of Michigan/CTIO Curtis Schmidt telescope, equipped with a SITE $2048 \times 2048$\ CCD, which gave a field of 1.35\degr\ at a scale of 2.3\arcsec\,pixel$^{-1}$ ($\sim$5~arcsec resolution). Both the LMC and SMC were mapped in narrow bands corresponding to \Halpha, \OIII\ ($\lambda$=5007\,\AA), and \SII\ ($\lambda$=6716,\,6731\,\AA). All the data have been flux-calibrated and assembled into mosaic images, a small section of which is shown in Fig.~\ref{xmmmcels}. Further details regarding the MCELS are given by \citet{2006NOAONL.85..6S} and at http://www.ctio.noao.edu/mcels. The MCELS data also reveal the SNR though at the poorer angular resolution it is hard to discern the full oval optical shell. However, MCELS does have the added advantage of the \OIII\ and \SII\ filters providing additional diagnostic capabilities. The composite RGB image with (R=H$\alpha$, G=[S\textsc{ii}], and B= [O\textsc{iii}]) centred on the new SNR is shown in Fig.~\ref{xmmmcels} with the \xmm\ contours in red (0.7--1.1~keV) and ATCA 5500~MHz radio data contours in white. Interestingly the strongest radio signature corresponds to the weakest optical emission to the south-east while the X-ray emission occupies the centre of the shell.

\subsubsection{AAT AAOmega spectroscopy}

\begin{table*}
\begin{center}
\caption{AAOmega fibre ID and fibre positions and line ratios for the spectral samples taken across the SNR as labelled in Fig.\ref{HAQ}. All line intensities are relative and extinction across the limited wavelength range of these emission lines is ignored. Errors in relative flux for observed lines are $\sim$10\%.}  
\label{fibrepos}
\begin{tabular}{ccccccccl}
\hline
\hline
\smallskip
2df fibre ID & RA (J2000) &DEC (J2000) & \SII/\Halpha & \Halpha\ & \SII6717 & \SII6731 & \SII6717/6731 & Comments\\ 
 \hline
Fib 238      & 05 08 16.2 & --69 03 34 & 0.62         & 938      & 372      & 209      & 1.8           & Low density limit\\
Fib 248      & 05 08 13.0 & --69 02 41 & 0.76         & 1241     & 638      & 301      & 2.2           & Low density limit\\
Fib 260      & 05 08 32.0 & --69 00 46 & 0.99         & 1248     & 640      & 594      & 1.2           & N$_{e}$ = 240\\
Fib 253      & 05 08 50.8 & --69 03 25 & 0.68         & 573      & 283      & 107      & 2.6           & Low density limit, low S/N\\
Fib 259      & 05 08 53.9 & --69 02 33 & 1.29         & 363      & 242      & 226      & 1.1           & N$_{e}$ = 395 low S/N\\
 \hline 
\end{tabular}
\end{center}
\end{table*}

As part of an AAOmega 2dF (two degree field) service night on the AAT on May 1st 2013, 2$\times$900 second exposures were taken on an LMC 2-degree field centred at RA(J2000)=05$^{h}$13$^{m}$15.7$^{s}$ and DEC(J2000)=--69$^{o}$29\arcmin51\arcsec\ for the LMC Planetary Nebula (PN) project of Reid and Parker \citep[e.g.][]{2006MNRAS.365..401R,2006MNRAS.373..521R,2013MNRAS.436..604R}. AAOmega is a powerful double beam spectrograph \citep{2006SPIE.6269E..14S} that is coupled to the 2dF multi-object fibre spectroscopy system on the AAT \citep{2002MNRAS.333..279L}. Due to the 400 available 2dF fibres it was possible to sacrifice five 2~arcsecond diameter fibres to allocate to the new SNR candidate that happened to be in the field of view of one of the Reid-Parker LMC PN fields. The individual SNR fibres were placed at carefully located positions on the western and eastern edges of the SNR shell as indicated in Fig.~\ref{HAQ} with their accurate positions listed in Table \ref{fibrepos}. 

Both the 580V and 1000R gratings were simultaneously employed for the general LMC-PN spectroscopic observations with the 1000R red grating centred at 6850\AA~ and giving coverage from 6300 to 7425\AA. This is ideal for medium resolution coverage of the diagnostic H$\alpha$, [NII] and [SII] nebular emission lines common to both PN and SNRs and also provides for more accurate radial velocity estimates. The lower resolution 580V blue grating was centred at 4800\AA~giving coverage
from 3700 to 5850\AA. The blue data suffered from the effects of extinction and low S/N and was of limited use for the SNR observations. 

\begin{figure}
\includegraphics[width=7.6cm,height=5.5cm]{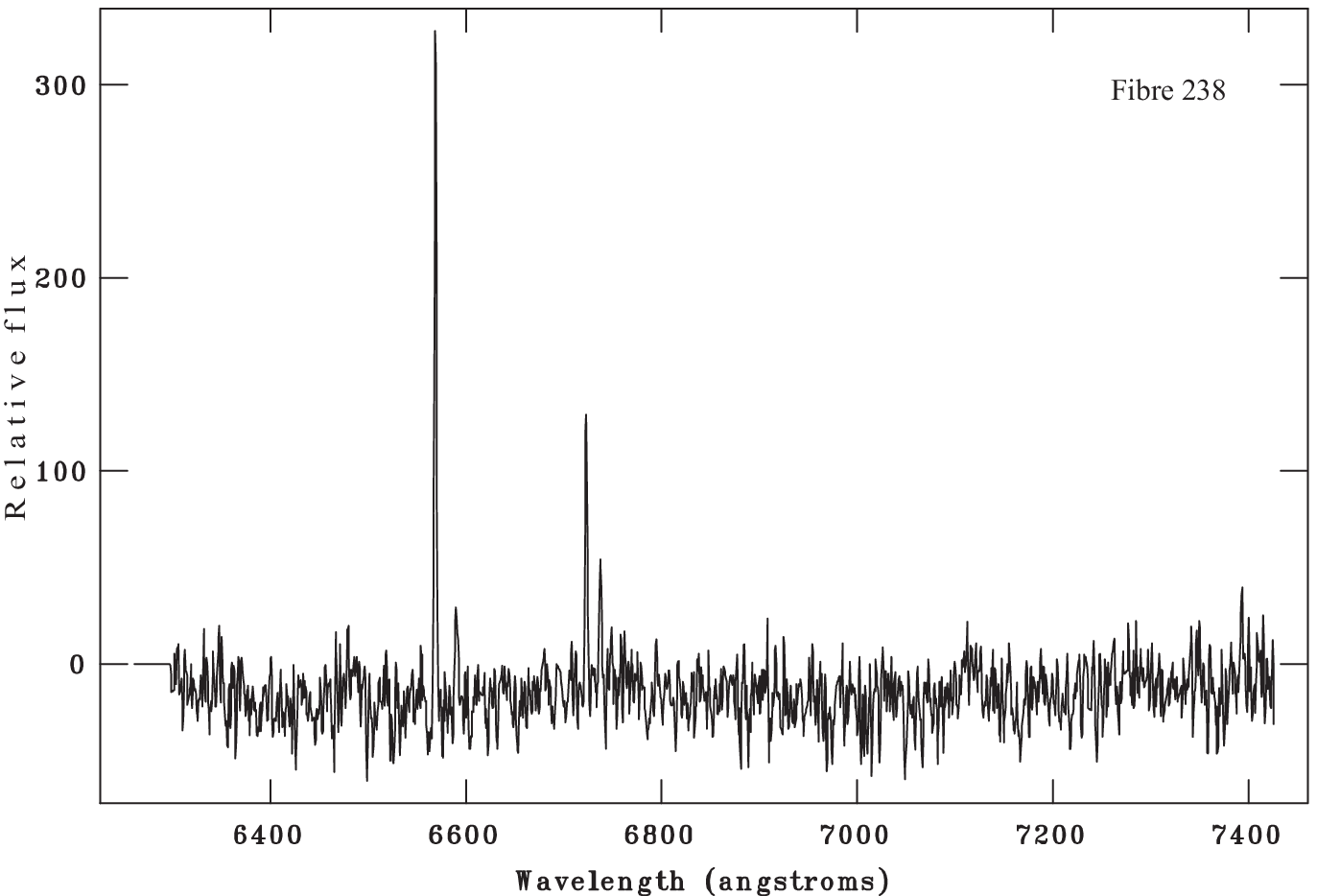}
\includegraphics[width=7.6cm,height=5.5cm]{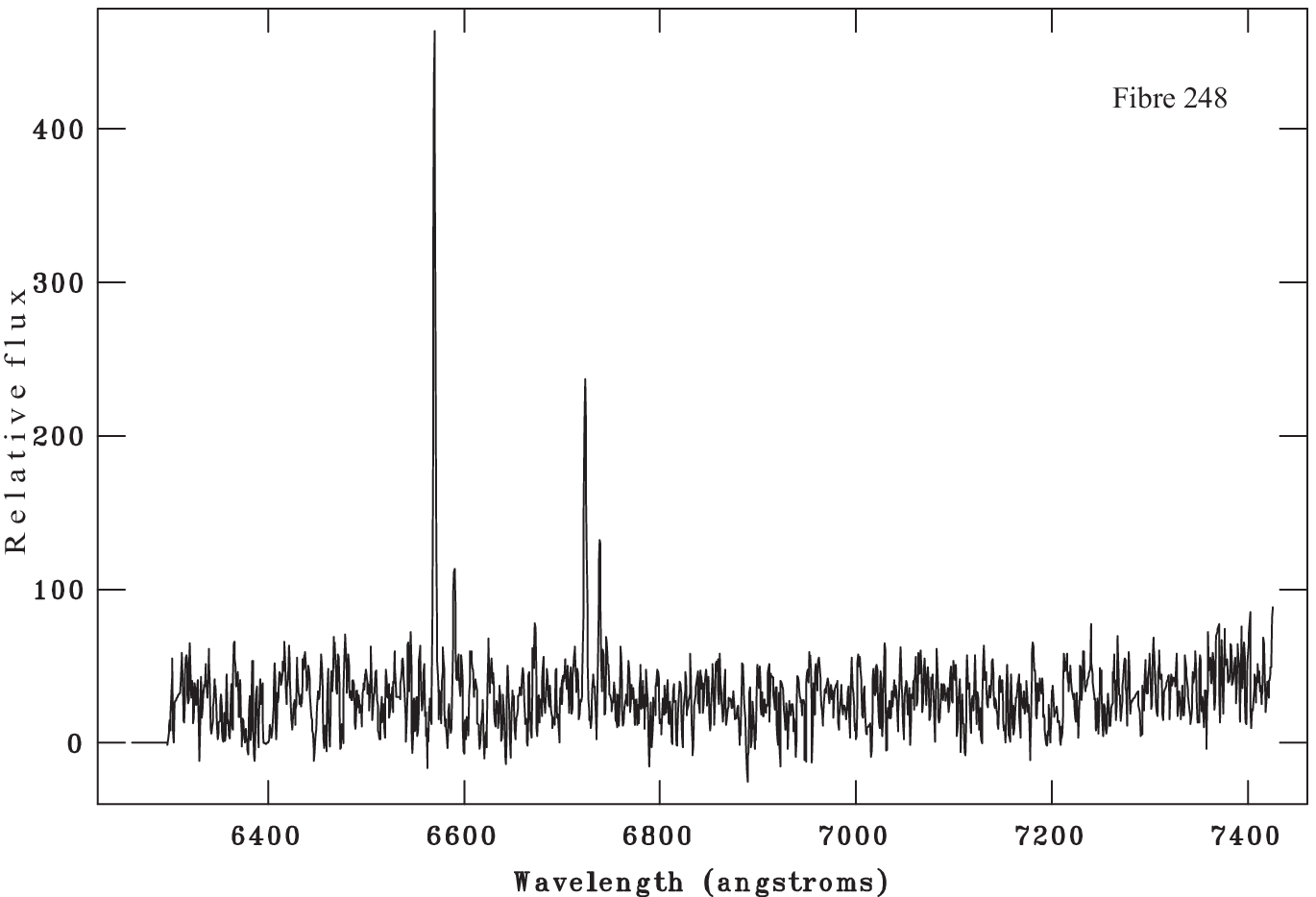}
\includegraphics[width=7.6cm,height=5.5cm]{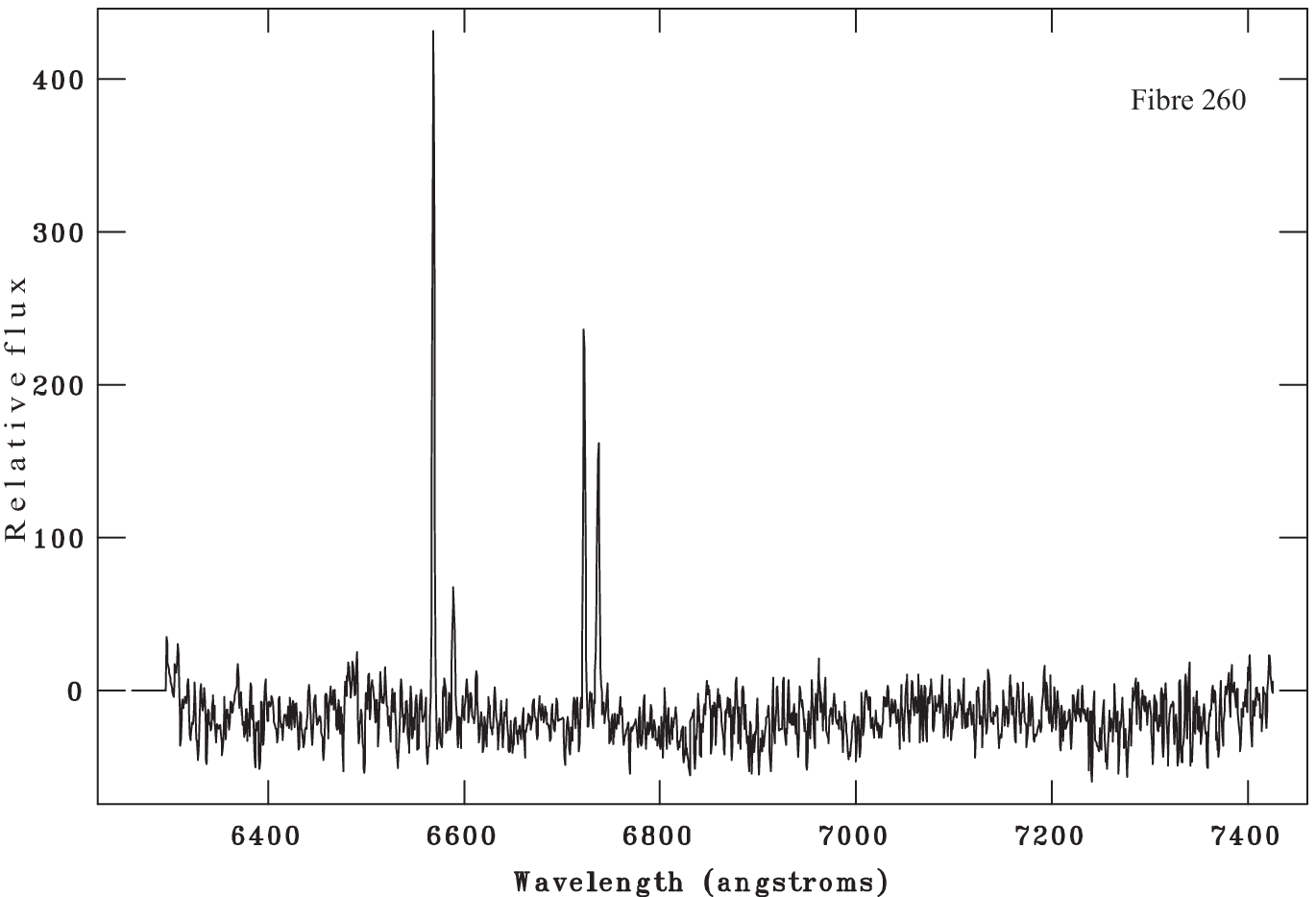}
\includegraphics[width=7.6cm,height=5.5cm]{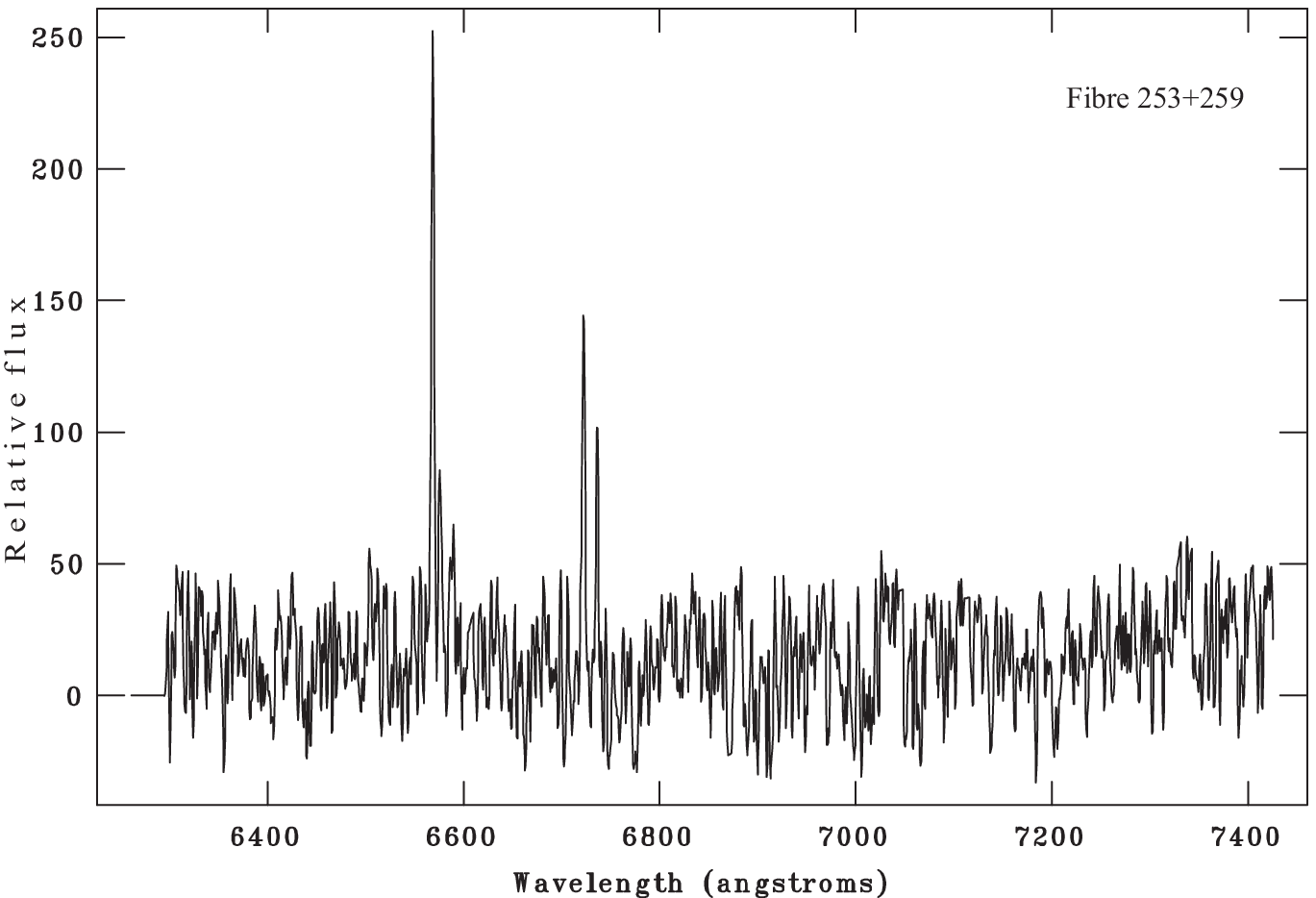}
\caption{1000R red grating 1-D 2dF spectra for Fibres 238, 248 and 260 taken on the enhanced western edge and a combined 1-D spectrum from Fibres 253 and 259 taken on the lower surface brightness eastern edge as indicated in Fig.~\ref{HAQ}.
 \label{2df-spec}}
\end{figure}

The AAOmega data was reduced with the standard 2dfdr data reduction pipeline (e.g. \citealt{2004AAONw.106...12C}). Note that the optical fibres sample both target and superposed sky. Hence it is standard practice to allocate a certain fraction of the 400 2dF fibres to blank sky. This is normally $\sim$20 fibres as was the case for these observations. However, because the LMC field is of very high stellar density, it was important to carefully manually select decent and representative sky positions across the 2-degree field to avoid stellar or nebulae contamination. An average of these sky spectra are then automatically subtracted from the individual target spectra as part of the 2dfdr data reduction process to provide the sky-subtracted spectra. The resultant 1-D fibre spectra taken at the various points labeled in Fig.~\ref{HAQ} are shown in Fig.~\ref{2df-spec}. The measured line ratios are provided in Table~\ref{fibrepos}. All line intensities are relative but the observed line-ratios, because they are so close in wavelength, will not be affected by the lack of a proper flux calibration (difficult for fibres) and so extinction across the limited wavelength range of these emission lines can also be safely ignored. 

It is a commonly adopted diagnostic \citep[see][]{1985ApJ...292...29F} that an [SII]/H$\alpha$ ratio $>$0.5 can separate SNRs from planetary nebula and HII regions as seen and expected in SNRs \citep[e.g][]{2012MNRAS.419.1413S}. It is clear from Table~\ref{fibrepos} that the observed [SII]/H$\alpha$ ratio which ranges from 0.6 to 1.29, is strongly in the shocked regime for all fibres, particularly for 2dF fibres 260 and 259 which also happen to be on opposing sides of the optical shell. These ratios all fall well within the SNR domain following the updated diagnostic emission line plots presented by \citet{2010PASA...27..129F}. Furthermore, the observed [SII]6717/6731\AA~line ratios also indicate the electron density is close to or in the low density limit as might be expected for such an evolved SNR. Nominal, unphysical, high values ($>$1.6) for the [S II]6717/6731\AA~ line ratios are observed in the lowest S/N spectra due to the associated larger measurement errors which are also indicated in Table~\ref{fibrepos}. 

The line intensities provided in Table~\ref{fibrepos} were measured using the IRAF $splot$ package. Gaussian profiles were fitted to individual lines, providing the central wavelength, the integrated line flux/intensity, equivalent width and Full Width at Half Maximum (FWHM). The IRAF $emsao$ package \citep{1995ASPC...77..496M} was used to determine the radial velocities for each object as presented in Table~\ref{fibreposv}, clearly showing that the object lies within the LMC. A small 2.5 km/s heliocentric velocity correction has been applied to the measurements. Errors in overall velocities are returned by $emsao$ after it estimates weighted averages for all lines measured. These can be seen to increase as the S/N decreases, particularly affecting fibres 259 and 253, whose combined spectra is shown in Fig.~\ref{2df-spec}. 

\begin{table}
\begin{center}
\caption{Heliocentric radial velocities estimated from the reduced 1-D 1000R red individual AAOmega fibre spectra as provided by the IRAF emsao package.}  
\label{fibreposv}
\begin{tabular}{cccl}
\hline
\hline
\smallskip
2df fibre ID & Radial velocity & Error & Comments\\ 
 & (km s$^{-1}$) & (km s$^{-1}$) & \\ 
 \hline
Fib 238      & 287.8 & 6.4 & for 2/5 lines\\
Fib 248      & 323.0 & 14.4 & for 5/5 lines\\
Fib 260      & 275.0 & 15.1 & for 3/5 lines \\
Fib 253      & 263.5 & 13.7 & for 3/6 lines\\
Fib 259      & 267.8 & 21.3 & for 5/6 lines \\
 \hline 
\end{tabular}
\end{center}
\end{table}

\section{Data Analysis}

\subsection{Radio}

The \textsc{miriad} \citep{1995ASPC...77..433S} and \textsc{karma} \citep{1995ASPC...77..144G} software packages were used for reduction and analysis. More information on the observing procedure and other sources observed in this session/project can be found in \citet{2012MNRAS.420.2588B,2012RMxAA..48...41B,2012SerAJ.185...25B,2013MNRAS.tmp.1294B} and \citet{2012A&A...540A..25D}. 

\begin{figure}
\centering\includegraphics[angle=-90,trim=0 50 0 0,scale=.37]{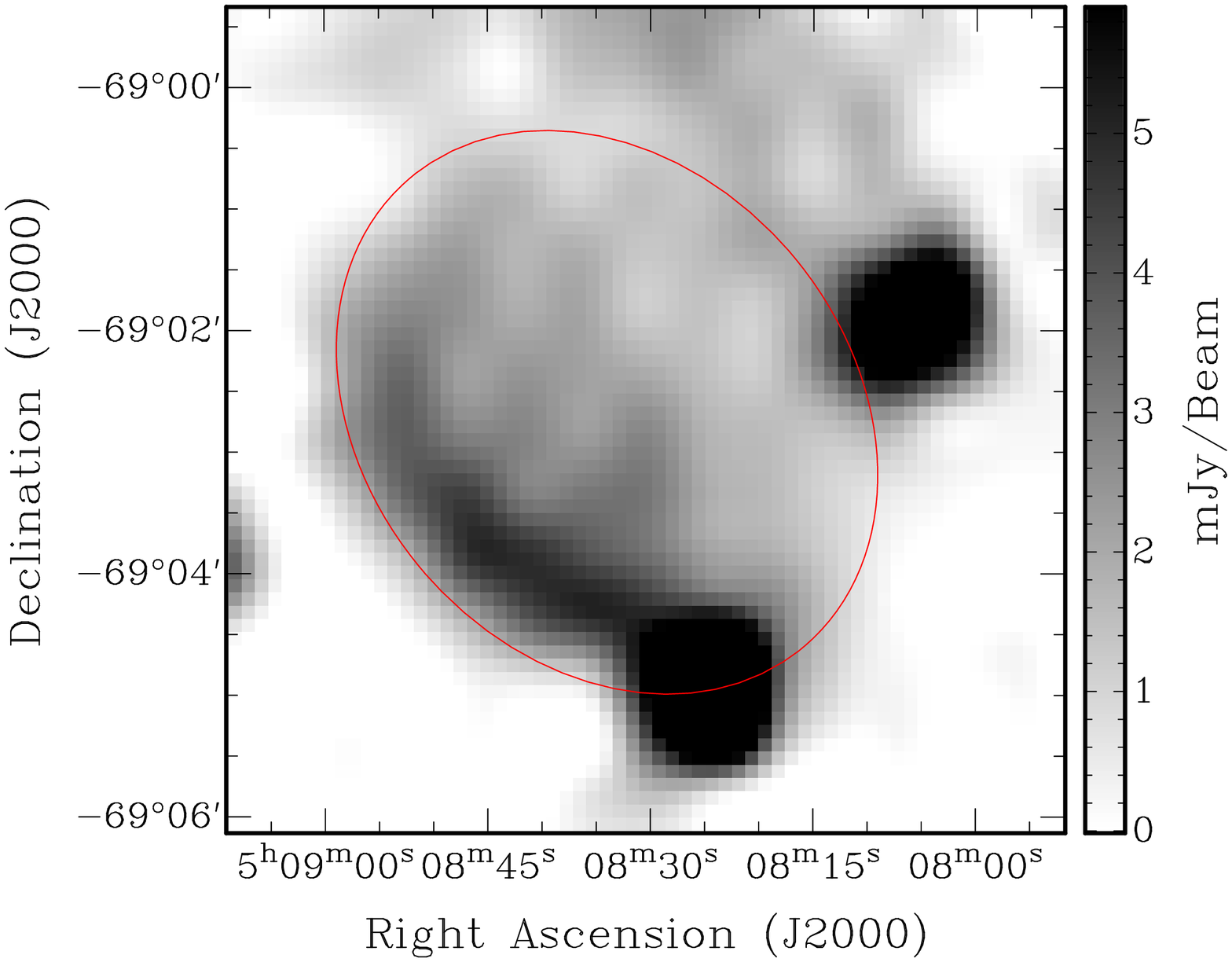}
\centering\includegraphics[angle=-90,trim=115 50 0 0,scale=.37]{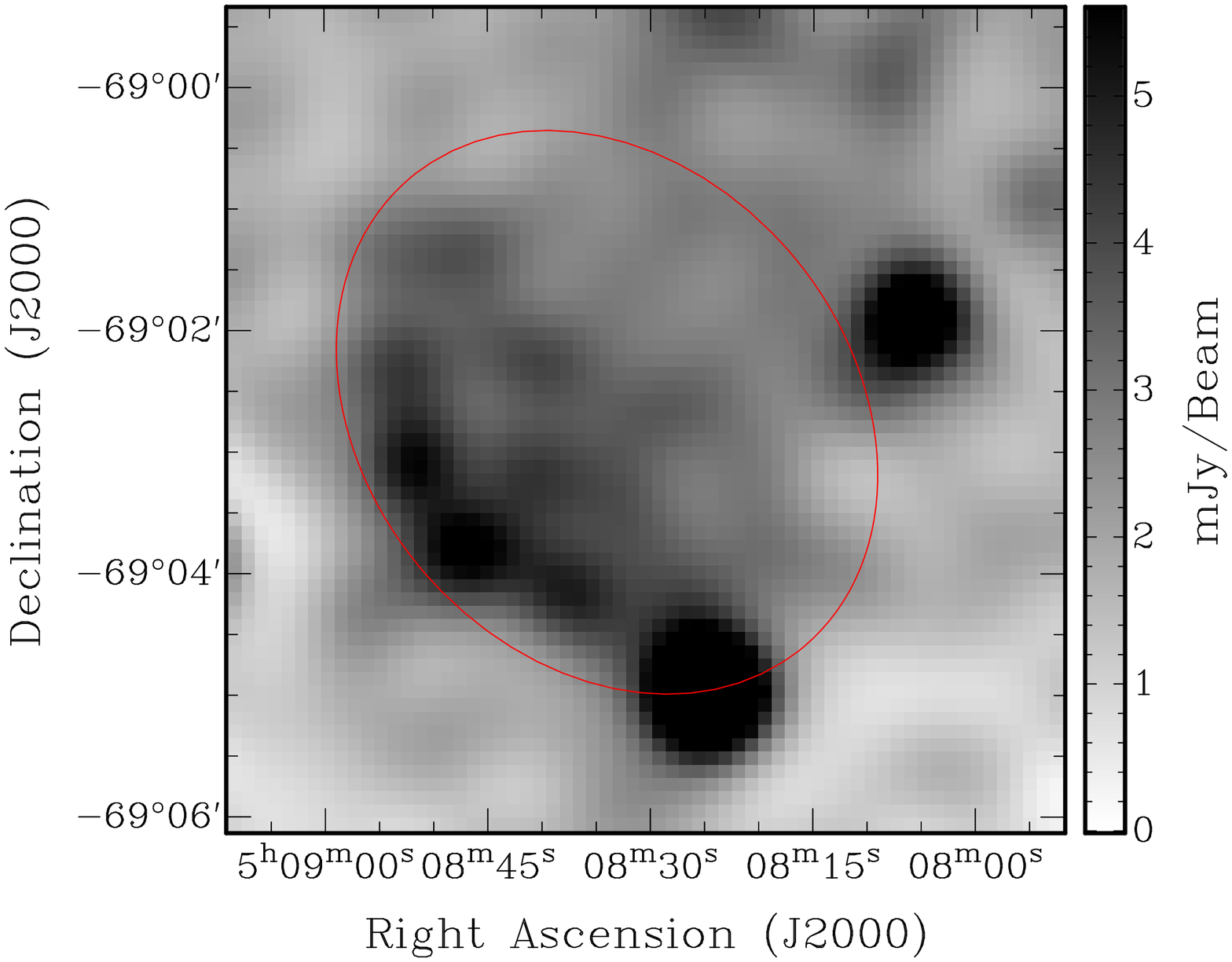}
\centering\includegraphics[angle=-90,trim=115 50 0 0,scale=.37]{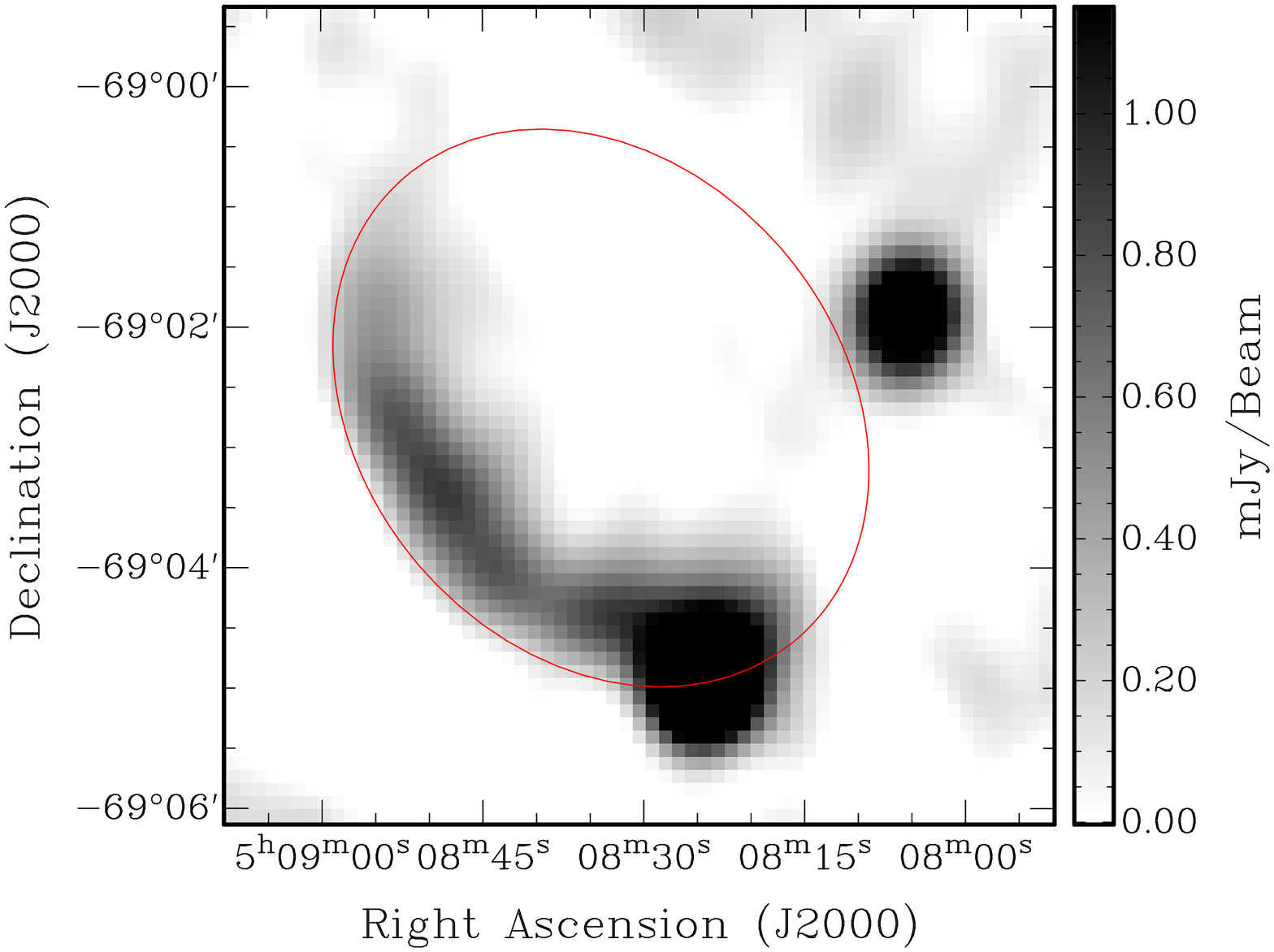}
\caption{843~MHz (top), 1400~MHz (middle) and 5500~MHz (bottom) radio-continuum images of \SNR\ with an overlaid ellipse indicating the approximate optical association for this SNR. The side bar quantifies the total intensity in mJy/beam).
 \label{rdimgs}}
\end{figure}

Images were formed using \textsc{miriad} multi-frequency synthesis \citep{1994A&AS..108..585S} and natural weighting. They were deconvolved using the {\sc mfclean} and {\sc restor} algorithms with primary beam correction applied using the {\sc linmos} task. The 6~cm (5500~MHz) ATCA image together with 20~cm (1337~MHz) and 36~cm (843~MHz) mosaic images of \SNR\ are shown in Fig.~\ref{rdimgs}. The resolution of the 6~cm image is 37.7\arcsec$\times$25.1\arcsec.

\subsection{X-rays}

\subsubsection{Imaging}
\label{x-ray_imaging}
We extracted images and exposure maps in three energy bands suited to the analysis of SNRs, whose X-ray emission is mostly thermal: a soft band (0.3--0.7~keV) including strong lines from oxygen; a medium band (0.7--1.1~keV), which includes Fe L-shell lines, as well as the He$\alpha$ and Ly$\alpha$ lines of Neon; and a hard band (1.1--4.2~keV) comprising lines from Mg, Si, S, Ca, Ar, and, if present, non-thermal continuum.

From the pn detector we selected single and double-pixel events (\texttt{PATTERN} = 0 to 4). Below 0.5 keV, only single-pixel events were selected to avoid the higher detector noise contribution from the double-pixel events.. All single to quadruple-pixel (\texttt{PATTERN} = 0 to 12) events from the MOS detectors were used. Bad pixels were masked. The detector background (see Sect. \ref{x-ray_background}) was taken from filter wheel closed (FWC) data\footnote{See \url{http://xmm2.esac.esa.int/external/xmm_sw_cal/background/filter_closed/index.shtml}}. Before it was subtracted from the raw images, we scaled it by a factor estimated from the count rates in the corners of the detectors not exposed to the sky.

We merged MOS and pn data, and combined images from the two \textit{XMM-Newton} observations. We then adaptively smoothed the images, computing Gaussian kernels at each position in order to reach an approximate signal-to-noise ratio of five. The minimum size for the kernels was set to a full width at half maximum (FWHM) of 20\arcsec. Smoothed images were finally divided by the vignetted exposure maps.

\begin{figure}
\centering\includegraphics[angle=-90,trim=0 0 0 0,scale=.4]{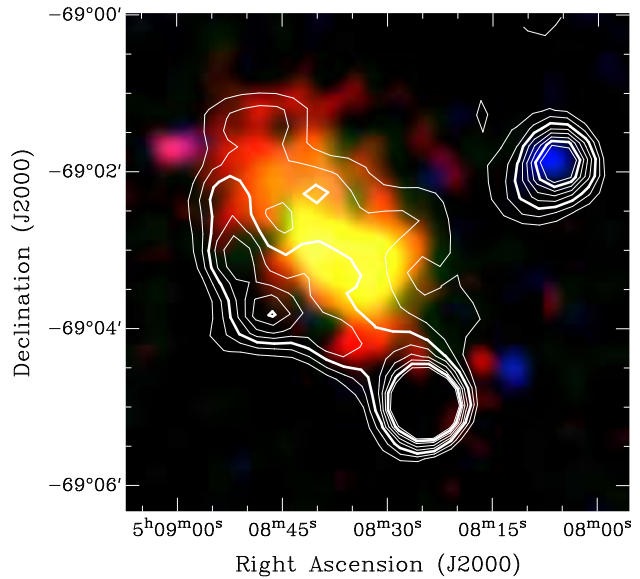}
\caption{\xmm\ three colour composite (red: 0.3--0.7~keV, green: 0.7--1.1~keV, blue: 1.1--4.2~keV) of \SNR\ smoothed with a Gaussian width of 20\arcsec, and overlaid with 1400~MHz radio contours. The contours are from 3$\sigma$ to 6.5$\sigma$ in steps of 0.5$\sigma$.
\label{xmmrad}}
\end{figure}

Combining the final images, we show in Fig. \ref{xmmrad} the \textit{XMM-Newton} `true-color' image of \SNR, which reveals a striking morphology. A bright central region, dominated by X-rays in the medium band, is surrounded by a fainter and softer shell. The brightness of the interior emission in the 0.7--1.1~keV band, which includes the Fe L-shell lines, suggests that \SNR\ falls into the class of SNRs with central Fe enhancements. Examples of these objects include 0548--70.4 and 0534--69.9 \citep{Hendrick2003}, DEM~L238 and DEM~L249 \citep{Borkowski2006}, and SNR~0104--72.3 (\citealt{Lee2011}, Roper et al., in preparation). Spectral differences between the central region and the shell are analysed in Section \ref{snr-x-ray}. There is little emission from the object above 1.1~keV. Several point sources are detected outside the boundary of the remnant, which are most likely background objects behind the LMC.

\subsubsection{X-ray spectral analysis}

We consider only Obs. ID 0690752001 for our spectral analysis. The $\sim$9 ks of data from Obs. ID 0651880201 can add
little to our analysis, as e.g. the faint shell of the remnant is not detected in images from this observation only. Furthermore, using the diagnostic devised by De Luca, Molendi, \& Leccardi \citeyearpar{2004A&A...419..837D} \footnote{See \url{http://xmm2.esac.esa.int/external/xmm_sw_cal/background/epic_scripts.shtml}}, we found that the flare--filtered event lists were still heavily contaminated by soft protons.

In addition, only EPIC-pn spectra are considered as the SNR shell is extremely faint in the EPIC-MOS images. Before proceeding with the spectral extraction we generated a vignetting-weighted event list to correct for the effective area variation across our extended source using the SAS task \textit{evigweight}. Only single and double-pixel events (\texttt{PATTERN} = 0 to 4) are considered. The extraction area of the central region was defined by enclosing the 0.7--1.1~keV band contours (see Fig. \ref{xmmmcels}) with an ellipse ($\sim 2.8\arcmin \times 1.6\arcmin$). The shell spectrum was extracted from a $\sim5\arcmin$ diameter region (as determined in Section \ref{morph}), excluding the central extraction region. Backgrounds should ideally be chosen from nearby positions on the detector. However, since the SNR was located close to on-axis and nearby suitable regions were comparatively source and chip-gap heavy, this was not possible. As such, backgrounds were extracted from a source and diffuse emission free region immediately to the northwest of the remnant, further off-axis than the source. This introduces important effects which must be considered in the spectral modelling of the particle-induced background, which is discussed in the forthcoming text. All spectra were rebinned using the FTOOL \textit{grppha}\footnote{See \url{http://heasarc.nasa.gov/lheasoft/ftools/fhelp/grppha.txt}} so that each bin contained a minimum of 30 counts. All fits were performed using XSPEC \citep{Arnaud1997} version 12.7.1 with abundance tables set to those of \citet{Wilms2000}, and photoelectric absorption cross-sections set to those of \citet{Bal1992}.

\subsubsection{X-ray background}
\label{x-ray_background}
Spectral analysis of extended X-ray sources is complicated by the contributions of the X-ray background, comprising particle-induced and astrophysical components. Often, a subtraction of a nearby background region is inappropriate due to the spectral and/or spatial variation of these background components. While the consideration of the X-ray background for bright extended sources may not be so significant, in the case of faint emission (as is the case with the shell of \SNR\ in particular), the background must be carefully accounted for in order to extract the purest possible information from the source. A commonly used method \citep[see][for a recent example]{2012A&A...546A.109M}, is to extract a nearby background spectrum, define a physical model for the background, and simultaneously fit this model to the source and background spectra. We adopt this method for our analysis of \SNR. A detailed description of our background treatment is given in Appendix A.\\

\subsubsection{\SNR\ emission}
\label{snr-x-ray-prep}
\par Line of sight absorption to \SNR\ consists of a Galactic component ($N_{\rm{H\ Gal}}$, fixed at $7\times10^{20}$ cm$^{-2}$ as described in Appendix \ref{A3}) and a foreground LMC component ($N_{\rm{H\ LMC,fg}}$). These were modelled using photoelectric absorption components with Galactic and LMC abundances, respectively.\\

\noindent \textit{Shell emission}
\par We modelled the shell emission using various thermal plasma models appropriate for SNRs. We report here on the results using the standard Sedov model \citep{Borkowski2001}, implemented in XSPEC as \textit{vsedov}. Given the large size of the remnant, the assumption that it is in the Sedov phase is justified. The free parameters of the \textit{vsedov} model are the mean shock temperature ($kT_{s}$), postshock electron temperature ($kT_{e}$), metal abundances, ionisation timescale ($\tau_{0}$, the electron density immediately behind the shock front multiplied by the age of the remnant). Due to the relatively poor count statistics, metal abundances for the model were fixed at the average LMC values (0.5~solar, \citealt{1992ApJ...384..508R}). In the case of \SNR, fixing the metal abundances at 0.5~solar is quite reasonable as we expect the shell to be dominated by swept-up ISM. \\

\noindent \textit{Interior emission}
\par Fits of the central spectrum are complicated by the fact that we must consider both the interior Fe-rich gas and the emission from the shell overlapping this region. To account for the shell emission we included a \textit{vsedov} component in the fits to the central spectrum with the shell $kT_{s}$ ($= kT_{e}$) and $\tau_{0}$ fixed to the best-fit values determined from fits to the shell spectrum (see above). For the interior emission itself we considered several thermal plasma models, which are discussed in detail in Section \ref{int-em}.

\section{Results}

\subsection{Radio}
This remnant exhibits a filled-in shell morphology (at 20~cm) centred at \mbox{RA(J2000)=05$^h$08$^m$33.7$^s$} and \mbox{DEC(J2000)=--69\degr02\arcmin33\arcsec}. There are two unrelated background sources in the field, one in the south (RA(J2000)=05$^h$07$^m$57.7$^s$, DEC(J2000)=--69\degr03\arcmin59.49\arcsec), and the other to the west (RA(J2000)=05$^h$08$^m$06.94$^s$, DEC(J2000)=--69\degr01\arcmin51.2\arcsec) (Fig.~\ref{rdimgs}). At lower radio frequencies in this study (843 \& 1400~MHz), the emission of the SNR is fairly uniform across the remnant with significant limb brightening in the south-east. As we move to higher frequencies ($\geq$5500~MHz), the remnant maintains this south-eastern limb but loses the shell emission of the remnant until we get to 9000~MHz, at which point all emission from the SNR is lost.

\begin{figure}
\centering\includegraphics[trim=0 0 0 0,scale=.55]{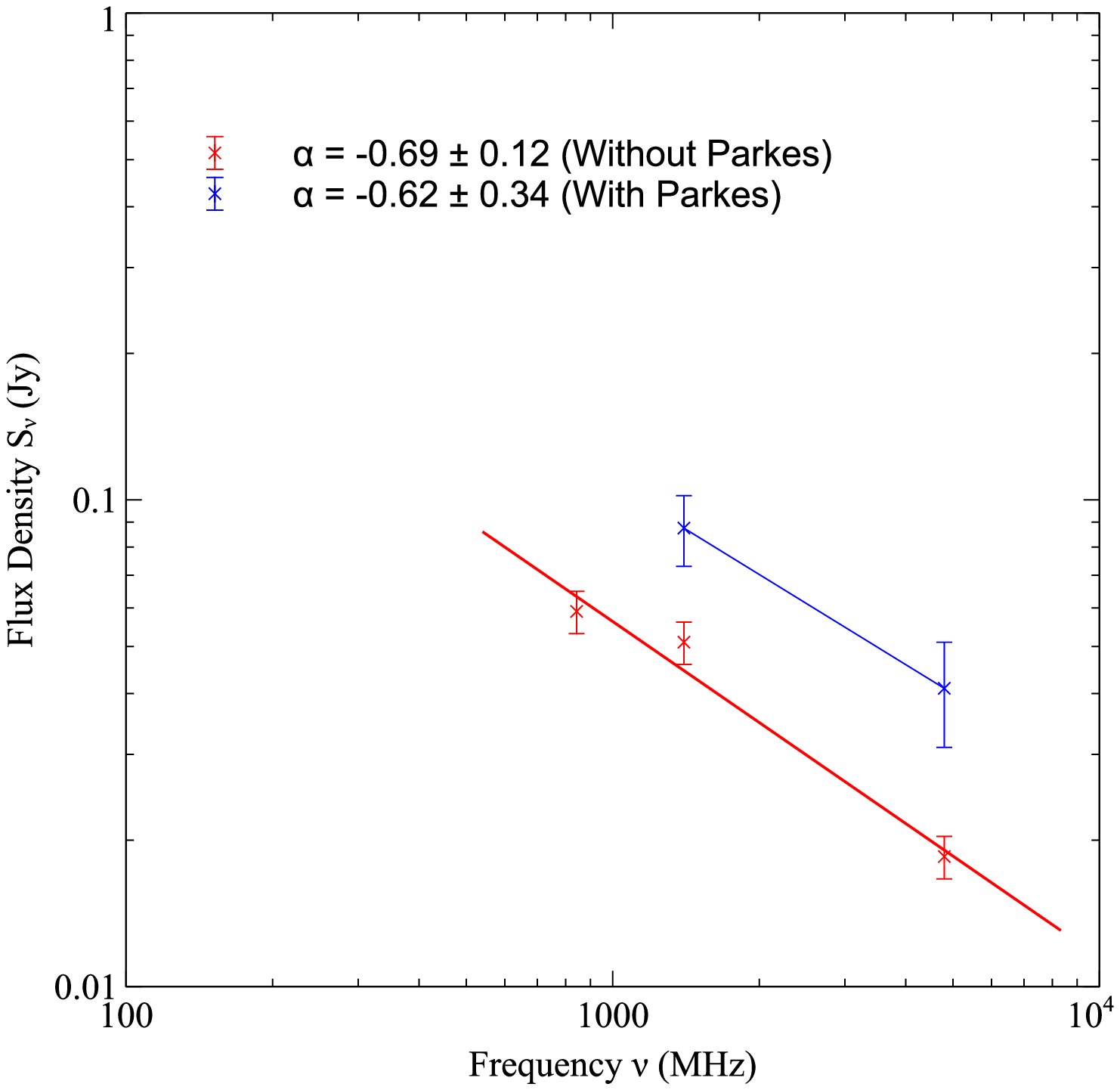}
\caption{Radio-continuum spectrum of \SNR.
 \label{spcidx}}
\end{figure}

The large extent of \SNR\ made a spectral energy distribution (SED) estimate difficult, as observations using an interferometer suffer from missing zero-spacings (which are responsible for large scale, extended emission), and therefore, missing flux. Because of this, we take two different sets of flux density measurements to estimate the SED. The first set of measurements came from mosaic images (20~cm from \citealt{2007MNRAS.382..543H} and 6~cm from \citealt{2010AJ....140.1567D}) which include a zero-spacing observation. The second set (which do not use a zero-spacing observation) include the two aforementioned mosaics before the zero-spacing (Parkes) observation were added, in addition to a 36~cm measurement from the Molonglo Synthesis Telescope (MOST) mosaic image (as described in \citealt{1984AuJPh..37..321M}). The values of these flux density measurements are reported in Table \ref{tbl-fluxes}, and then used to produce a spectral index graph (Fig.~\ref{spcidx}). 
The spectral indices for both the zero-spacing inclusive ($\alpha=-0.62\pm0.34$) and the non inclusive ($\alpha=0.69\pm0.12$) are quite similar. We note however, as \textit{u--v} coverage is a function of distance and frequency, the 6~cm measurement without the short-spacing is missing more extended emission than those at lower frequencies. Therefore, we would expect a flatter spectral index than we see from these measurements (and potentially closer to what we see in the zero-spacing measurements).

\begin{table}
 \caption{Integrated flux densities of \SNR\ measured in this study.}
  \begin{threeparttable}
 \label{tbl-fluxes}
 \begin{tabular}{@{}cccccccl}
  \hline
$\lambda$      & $\nu$  & R.M.S  & Beam Size & S$_\mathrm{Total}$ & $\Delta$S$_\mathrm{Total}$ \\
(cm) & (MHz)  & (mJy) & (\arcsec) & (mJy) & (mJy)    \\
\hline
20 & 1377 & 0.7 & 40.0$\times$40.0 & 88 & 15 \\
6 & 4800 & 1.0 & 35.0$\times$35.0 & 41 & 10 \\ 
\hline
36 & 843 & 0.4 & 46.4$\times$43.0 & 59 & 6 \\
20 & 1377 & 0.7 & 40.0$\times$40.0 & 51 & 5 \\
6 & 4800 & --- & 33.0$\times$33.0 & 19 & 2 \\ 
\hline
 \end{tabular}
 \begin{tablenotes}
\item The top two flux density measurements include a zero-spacing measurement, whereas the lower three measurements do not.
 \end{tablenotes}
 \end{threeparttable}
\end{table}

There was no reliable detection in the {\it Q} or {\it U} intensity parameters associated with this object, implying a lack of polarisation at higher radio frequencies (5500 \& 9000~MHz). Without these detections, we were unable to determine the Faraday rotation and therefore cannot deduce the magnetic field strength. As an alternative however, we were able to estimate the magnetic field strength for this SNR based on the equipartition formula as given in \citet{2012ApJ...746...79A}. This formula is based on the \citet{1978MNRAS.182..443B} diffuse shock acceleration (DSA) theory. We use the spectral index which is inclusive of zero-spacing measurements ($\alpha = -0.62$) for this equipartition measurement. We find the average equipartition field over the whole shell of SNR \SNR\ to be $\sim$28~$\mu$G (this value is typical for the compressed interstellar magnetic field by the strong shock wave of an older SNR), and an estimated minimum SN explosion energy of E$_{min}$ = 1.6$\times$10$^{50}$ ergs (see \citet{2012ApJ...746...79A}; and corresponding calculator\footnote{The calculator is available on \url{http://poincare.matf.bg.ac.rs/~arbo/eqp/ }}).

From the position of \SNR\ at the surface brightness to diameter ($\Sigma$ - D) diagram ((D, $\Sigma$) = (65.5~pc, 1.4~$\times$~10$^{-21}$~W~m$^{-2}$~Hz$^{-1}$~sr$^{-1}$)) by \citet{2004A&A...427..525B}, we can estimate that \SNR\ is likely to be an older SNR probably in the  transition phase between Sedov and radiative phases of evolution.

\subsection{Multifrequency Morphology}
\label{morph}
The morphology of the soft X-ray emission (0.3--0.7~keV) of \SNR\ is similar to that of the 20~cm radio-continuum emission (Fig.~\ref{xmmrad}), filling out the shell of the remnant. X-ray emission in the medium band (0.7--1.1~keV) appears to be encased within the radio and optical emission, making the central location of this emission evident.

In the MCELS combined RGB \Halpha, \OIII\ and \SII\ optical image we can see an association between the \OIII\ emission and radio emission (Fig.~\ref{xmmmcels}) where the \OIII\ emission perfectly fits within the confines of the radio emission along the south-eastern limb of the remnant and then extending in the northern region to further complete the ellipse ring structure of the remnant. On the other hand, the \Halpha\  and \SII\ emission located in the north-western regions of the remnant show no association with the radio-continuum emission at any frequency, nor with the X-ray emission (Fig.~\ref{xmmmcels}). The \Halpha\ and \SII\ emission in the north-west is significantly stronger when compared to the faint \OIII\ emission. Strangely enough, they almost perfectly complement the \OIII\ emission in creating an ellipse ring morphology. This morphology is similar to that of LMC SNR J0453--6829 \citep{2012A&A...543A.154H}.

\begin{figure}
\centering\includegraphics[trim=40 0 0 70,scale=.58]{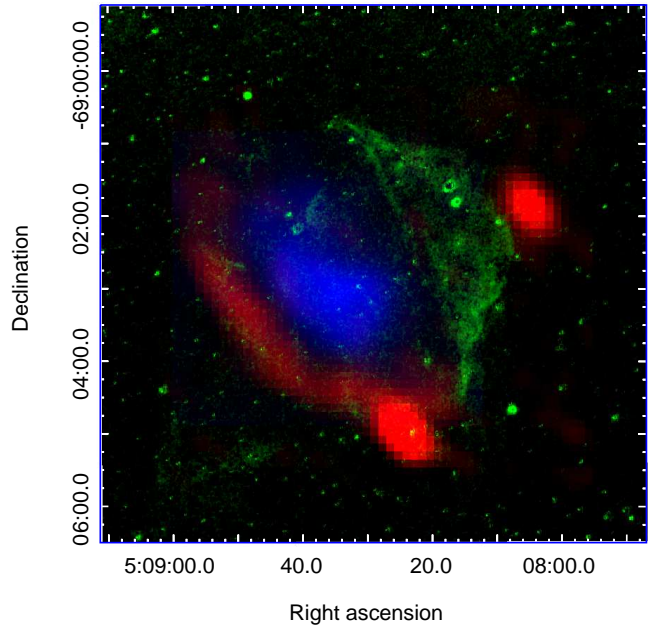}
\caption{Radio (5500~MHz) [red], Optical (\Halpha) [green] \& X-ray (0.7--1.1~keV) [blue] emission of \SNR.
 \label{comp}}
\end{figure}

Due to the higher resolution of \Halpha\ images from the deep \Halpha\ survey of the central 25~sq. degrees of the LMC by \citet{2006MNRAS.365..401R} (Fig.~\ref{HAQ} \& \ref{comp}), we are able to see a complete optical shell which perfectly follows the radio emission in the  south-east. The enhanced optical emission seen in this image towards the north-west suggests propagation of the shock front into the slightly higher density ISM, compared with somewhat lower density ISM on the south-east side. Separate fragmented \Halpha\ filaments inside the shell are common in optical observations (as well as in radio), as \Halpha\ emission is dominant where the blast propagates into dense, cooler regions. 

X-ray, radio and optical images were combined to estimate the extent of this object. To do so, we created an annotation file which comprised of an ellipse that followed the outer emission from these combined images. We deduce an extent of 304\arcsec$\times$234\arcsec$\pm$4\arcsec\ (74$\times$57$\pm$1~pc), which places this remnant amongst the 20\% largest SNRs in the LMC \citep{2010MNRAS.407.1301B}.

\begin{figure}
\centering\includegraphics[trim=0 25 0 0, angle=-90, scale=.45]{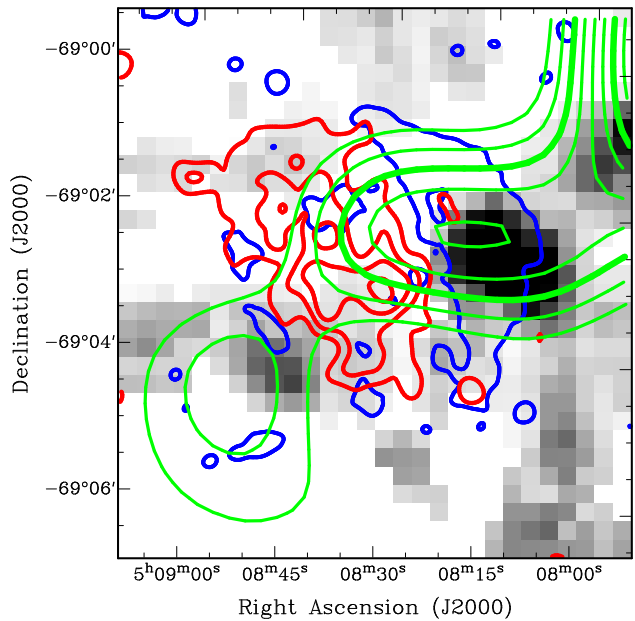}
\caption{\xmm\ (0.3--0.7~keV; red) and \Halpha\ (blue) contours of \SNR\  overlaid on NANTEN molecular cloud contours (green) and MAGMA molecular cloud grayscale intensity image, showing the association between the denser environment and the emission from the remnant.
\label{molecxmm}}
\end{figure}

\begin{figure*}
\centering\includegraphics[trim=1cm 8cm 0.5cm 0.5cm, clip=true, angle=0, width=0.99\linewidth]{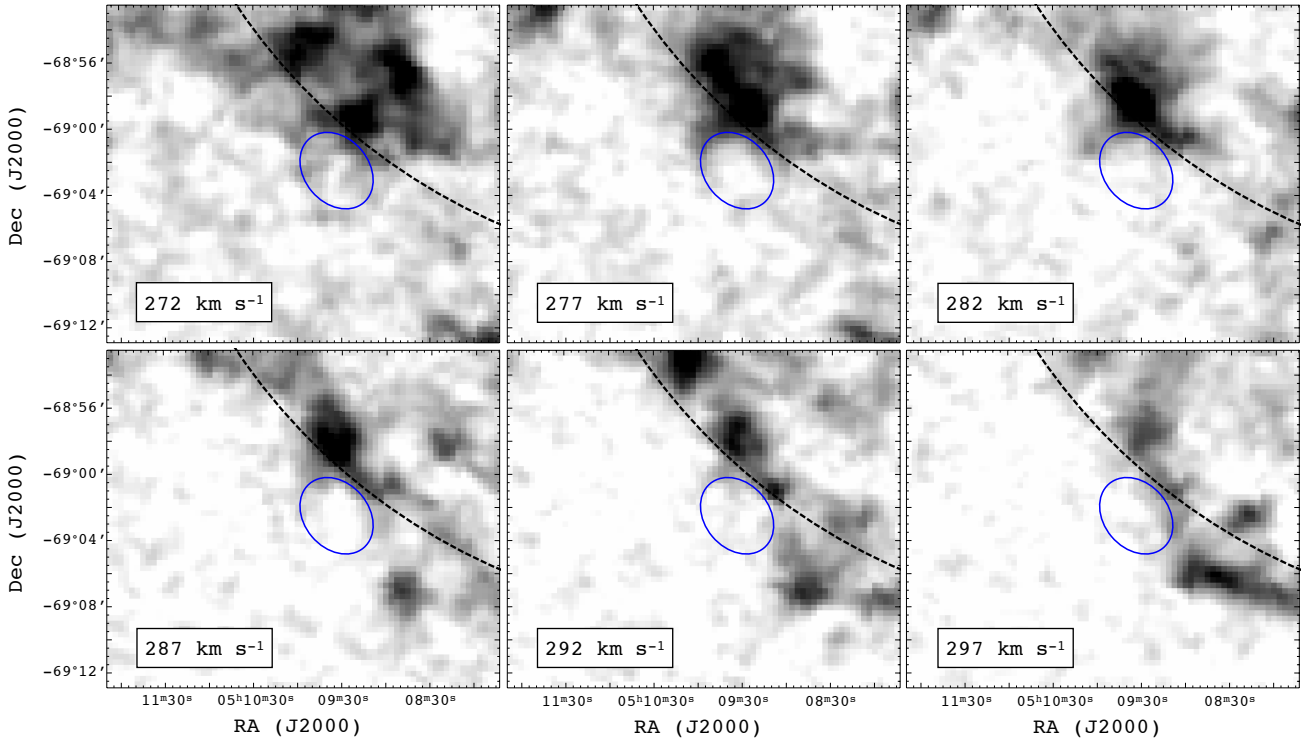}
\caption{HI channel maps of SGS~5 in the region of \SNR. The dimensions of the SNR are shown by the blue ellipses. The black dashed lines mark the approximate SGS~5 dimensions as determined by \citet{1999AJ....118.2797K}. The heliocentric velocities are also indicated. A density gradient across \SNR\ due to SGS~5 is evident in each channel.}
\label{maps}
\end{figure*}

To gain an understanding of the environment in which the remnant resides, we use an image from the NANTEN molecular cloud survey by \citet{2009ApJS..184....1K}, and another from the higher resolution Magellanic Mopra Assessment (MAGMA) survey by \citet{2011ApJS..197...16W}. From these images (Fig. \ref{molecxmm}), it appears that the remnant is confined within the CO emission, elongated and stretching into the less dense region towards the north-east. However, this may be a projection effect as there is no reason to believe that the SNR is at the same distance as the molecular cloud. If it is though, perhaps the expansion could have been stopped by the cloud but some of it got broken up and ionized by the radiation from the explosion, giving rise to the asymmetry. \SNR\ was also found to be located at the rim of the supergiant HI shell SGS~5, identified by \citet{1999AJ....118.2797K}. We recreated their HI channel maps of SGS 5 (their Fig. 5e) at the position of \SNR\ using ATCA-Parkes 21 cm observations from the HI Magellanic Cloud Survey\footnote{Available at \url{http://www.atnf.csiro.au/research/HI/mc/}}, shown in Fig. \ref{maps}. The ATCA and Parkes observations are outlined in \citet{1998ApJ...503..674K} and \citet{2003ApJS..148..473K}, respectively. By definition, HI supergiant shells have radii that exceed the neutral gas scale height of the LMC ($\sim$180~pc) and, thus, have blown out both sides of the disk, effectively creating a cylindrical wall of HI gas. Indeed, the molecular cloud shown in Fig. \ref{molecxmm} is most likely located, and may have been formed in SGS 5 \citep{2013ApJ...763...56D}. The extent of SGS~5 through the LMC disk, increases the likelihood of an interaction with \SNR\ and can provide the density gradient required to explain the asymmetric optical and radio emission.

Because of the bright interior X-ray emission, this remnant could be suggested to be a member of the mixed--morphology SNR (MMSNR, \citealt{1998ApJ...503L.167R,2006ApJ...647..350L}). These remnants shows indications for interactions with molecular clouds and often show associated OH maser emission. These features led some authors to conclude to a core-collapse origin for these type of SNRs (e.g. Micelli 2011). In the case of \SNR, although there is some evidence of interaction with a molecular cloud, we clearly favour a type Ia SN progenitor (see Sect. 4.3.2). We note that other type Ia SNRs have been found in a molecular cloud/star forming environment (e.g. \citealt{2003ApJ...582..770L, 2011ApJ...731L...8L}). In addition, over-ionised (i.e. recombining) plasma have been found recently in several Galactic MMSNRs \citep{2009ApJ...705L...6Y, 2011PASJ...63..527O, 2013PASJ...65....6Y}, while this feature is absent in the spectrum of MCSNR J0508-6902. Finally, we note that this classification has only been applied to Galactic remnants and not to the (usually Fe-rich) `centrally-peaked in X-rays' SNRs in the Magellanic Clouds (see Sect. 5 and references therein).

\subsection{X-ray emission of \SNR}
\label{snr-x-ray}
\subsubsection{Shell emission}
As stated in Section \ref{snr-x-ray-prep}, we modelled $N_{\rm{H\ LMC,fg}}$ using a photoelectric absorption model with LMC abundances. However, $N_{\rm{H\ LMC,fg}}$ consistently tended to zero, with upper 90\% confidence intervals for this parameter of \mbox{$\sim8\times10^{20}$ cm$^{-2}$} for all fits. Thus, we fixed $N_{\rm{H\ LMC,fg}}$ at 0. Initial shell fits with the \textit{vsedov} model resulted in best-fit values for $kT_{s}$ and $kT_{e}$ to be equal within their 90\% confidence intervals and, thus, $kT_{s}$ and $kT_{e}$ were constrained to be the same. Indeed, we should expect that evolved SNRs are close to ion-electron temperature equilibrium \citep{Borkowski2001}. \\

The results of the spectral fits to the shell and central regions of \SNR\ are given in Table \ref{fit-results}. The \textit{vsedov} model provides an acceptable fit to the shell spectrum with a reduced $\chi^{2}$ of 1.21. The best-fit temperature to the shell emission of $kT_{s} =  kT_{e} = 0.41(^{+0.05}_{-0.06})$~keV is consistent with other large LMC SNRs \citep[see][for example]{Williams2004,Grondin2012}, as well as DEM~L238 and DEM~L249 from \citet{Borkowski2006}. The high value of $\tau_{0}$ ($\sim 10^{12}\ \rm{s\ cm}^{-3}$) indicates that the plasma is in ionisation equilibrium. From the shell component in the fits to the central spectra, we determined that 13-28\% of the \textit{total} shell emission is contributing to the central spectrum. The absorption corrected X-ray luminosity ($L_{\rm{X}}$) of the shell in the 0.2--10~keV range was found to be $\sim5\times10^{34}\ \rm{erg\ s}^{-1}$.\\

\begin{table}
\caption{Results of the spectral fits. For the \vsedov\ model, the postshock mean temperature and postshock electron temperature were set to be the same ($kT_{s} =  kT_{e} = kT$) and the abundance is set to that of the LMC (0.5 solar). The EM and  $L_{\rm{X}, 0.2-10\ \rm{keV}}$ values for the shell were calculated for the \textit{total} shell (i.e., central and shell region). The ranges in parentheses represent the 90\% confidence intervals on the parameters.}
\begin{center}
\begin{normalsize}
\label{fit-results}
\begin{tabular}{p{7.5cm}}
\hline
\hline
Shell emission: $vsedov$ model \\
Fit parameter/result \hfill Value \\
\hline
$kT$ (keV)\dotfill0.41 (0.35--0.46) \\
$\tau$ ($10^{12}$ s cm$^{-3}$)\dotfill2.70 (1.73--6.92)  \\
$\chi^{2}/\nu$\dotfill267.08/219 \\
$EM$ ($10^{57}$ cm$^{-3}$)\dotfill4.75 (2.54--7.22) \\
$L_{\rm{X}, 0.2-10\ \rm{keV}}$ ($10^{34}$ erg s$^{-1}$)\dotfill5.1\\
\hline
\hline
Interior emission: pure heavy element $vapec$ models \\
Fit parameter \hfill Value \\
\hline
$kT_{\rm{Fe, O}}$ (keV)\dotfill0.78 (0.75--0.81) \\
$\chi^{2}/\nu$\dotfill66.49/64\\
$n_{e}n_{Fe}V$ ($10^{53}$ cm$^{-3}$)\dotfill0.58 (0.54--0.61)\\
$n_{e}n_{O}V$ ($10^{53}$ cm$^{-3}$)\dotfill2.99 (0--9.96)\\
$L_{\rm{X}, 0.2-10\ \rm{keV}}$ ($10^{34}$ erg s$^{-1}$)\dotfill2.8\\
\hline
\end{tabular}
\end{normalsize}
\end{center}
\end{table}%

\begin{figure*}
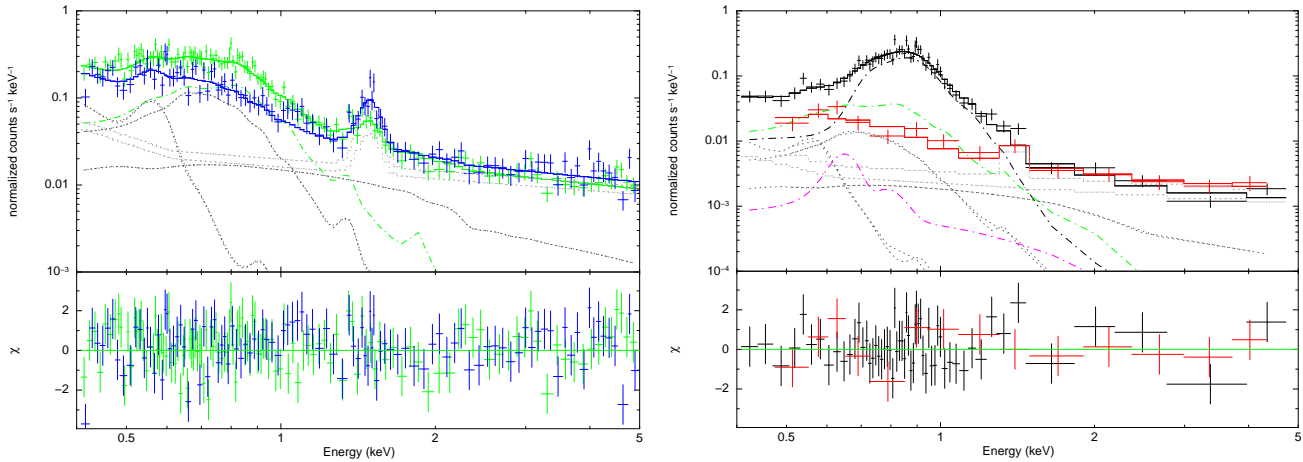

\centering
\begin{subfigure}
  \centering
  \includegraphics[trim= 1.2cm 0.9cm 0.5cm 2.9cm, clip=true, angle=270,width=.49\linewidth]{shell}
\end{subfigure}%
\begin{subfigure}
  \centering
  \includegraphics[trim= 1.2cm 0.9cm 0.5cm 2.9cm, clip=true, angle=270,width=.49\linewidth]{central}
\end{subfigure}
\caption{{\it XMM-Newton} spectral fits to \SNR\ shell (left) and central spectra (right) and associated backgrounds. The shell spectrum is shown in green, the shell background in blue, the central spectrum in black, and the central background in red, with the best-fit model to each represented by the solid lines of matching colour. The astrophysical background components are indicated by the dark grey dotted lines in each plot. The particle induced background components are indicated by the light grey dotted lines. The green dash-dot line in each spectrum represents emission from the shell. The black and magenta dash-dot lines in the central spectrum show the pure Fe component and the pure O component, respectively.}
\label{x-ray-spectrum}
\end{figure*}

Using our fit results we can further estimate physical parameters of the SNR using the Sedov dynamical model \citep[see][for example]{Sasaki2004}. We know that \SNR\ is not spherically symmetric since the semi-major and semi-minor axes are 37 pc and 28.5 pc, respectively.  Taking these semi-major and semi-minor axes as the first and second semi-principal axes of an ellipsoid describing \SNR, and assuming that the third semi-principle axis is in the range 28.5-37 pc, we determined the volume ($V$) limits for the remnant and their corresponding effective radii ($R_{\rm{eff}}$) to be $(3.7-4.8)\times10^{60}$ cm$^{3}$ and 31.1--33.9 pc, respectively.  In the following calculations we evaluate the physical parameters of the remnant at these volume and effective radii limits, and determine the resulting range for each parameter.\\

\par The X-ray temperature corresponds to a shock velocity

\begin{equation}
v_{s}=\left(\frac{16kT_{s}}{3\mu}\right)^{0.5},
\end{equation}

where $kT_{s}$ is the postshock temperature and $\mu$ is the mean mass per free particle. For a fully ionised plasma with LMC abundances, $\mu=0.61$, resulting in a shock velocity of $v_{s}~=~572(\pm36)$~km~s$^{-1}$. The age of the remnant can now be determined from the similarity solution:

\begin{equation}
v_{s}=\frac{2R}{5t},
\end{equation}

where $R=R_{\rm{eff}}$ and $t$ is the age of the remnant. This gives an age range of $20-25$~kyr. The preshock H density ($n_{\rm{H},0}$) in front of the blast wave can be determined from the emission measure ($EM$, see Table \ref{fit-results}). Evaluating the emission integral over the Sedov solution using the approximation for the radial density distribution of \citet{Kahn1975} gives

\begin{equation}
\int n_{e}n_{\rm{H}}dV=EM=2.07\left(\frac{n_{e}}{n_{\rm{H}}}\right)n_{\rm{H},0}^{2}V,
\end{equation}

where $n_{\rm{e}}$ and $n_{\rm{H}}$ are electron and hydrogen densities, respectively, V is the volume \citep[e.g.,][]{1983ApJS...51..115H}. Taking $n_{e}/n_{\rm{H}}=1.21$ and the determined range of volumes, this equation yields $n_{\rm{H},0}=(1.5-2.8)\times10^{-2}$~cm$^{-3}$. Since the preshock density of nuclei is given as $n_{0}\sim1.1 n_{\rm{H},0}$, it follows that  $n_{0} = (1.7-3.1)\times10^{-2}$~cm$^{-3}$. Hence, the SNR is expanding into a fairly rarified environment. The initial explosion energy ($E_{0}$) can be determined from the equation:

\begin{equation}
R=\left(\frac{2.02E_{0}t^{2}}{\mu_{n}n_{0}}\right)^{1/5},
\end{equation}

where $\mu_{n}$ is the mean mass per nucleus ($=1.4m_{p}$). This results in an initial explosion energy in the range $(0.37-0.75)\times10^{51}$~erg which is less than the canonical value of $10^{51}$~erg but consistent with several Type Ia SNRs in the LMC \citep[see][for example]{Hendrick2003}. The swept-up mass contained in the shell is given simply by $M=V\mu_{n}n_{0}$, which is evaluated to ($32-62$)~M$_{\odot}$. This large amount of swept-up material supports our original assumption that \SNR\ is into the Sedov phase of its evolution. All the derived parameters of \SNR\ are summarised in Table \ref{physical-properties}. Using the determined values of $t$ and $n_{0}$ we estimated the expected ionisation age for the remnant, given as $4.8n_{\rm{H},0}t$, to be in the range $(5.3-9.2)\times10^{10}$~cm$^{-3}$ s. This is significantly less than the ionisation age determined from the spectral fits with the Sedov model. Given this inconsistency, it is likely that the Sedov model may not be strictly applicable to \SNR. Several other remnants exhibit a similar discrepancy between measured and expected ionisation timescales. \citet{1998ApJ...505..732H} and \citet{2003A&A...406..141V} observed this for candidate CC SNRs in the LMC and SMC, respectively. \citet{1998ApJ...505..732H} suggest the explanation to be that the remnants initially expand into a low density cavity in the ISM, likely due to the massive stellar wind. The remnant rapidly reaches the wall of the cavity where it meets the denser material, slows down, and emits X-rays. This scenario represents a deviation from the standard Sedov model of a remnant expanding into a uniform ambient medium and can account for the discrepancy in the measured and expected ionisation ages. The problem with this scenario as an explanation for \SNR\ is that a high mass stellar wind is required to alter the ambient ISM in such a manner which is likely not the case for \SNR. \citet{Borkowski2006} also observe such a discrepancy for DEM L238, a Type Ia remnant. They suggest that their derived ambient ISM density may be underestimated resulting in the ionisation parameter discrepancy. It is unclear as to the reason for the observed discrepancy in \SNR, however, it is most likely the case that it is due to some density variation in the ambient ISM causing a deviation from the standard Sedov picture. The optical and radio morphologies discussed in Section \ref{morph} suggest that \SNR\ is evolving into a denser medium in the west. The higher $n_{\rm{H},0}$ value in the west of the remnant will result in an increased ionisation age in the west as compared to the east. This may explain the observed ionisation parameter discrepancy. However, in such a picture we would expect variation of the X-ray emission in this direction. Unfortunately, given the already low count statistics, an analysis of spatio-spectral variation is not really feasible. In any case, because of the discrepancy between the measured and expected ionisation parameters, the SNR properties derived from the Sedov dynamical model are somewhat uncertain.

\begin{table}
\caption{Physical properties of \SNR\ derived from the Sedov model}
\begin{center}
\begin{normalsize}
\label{physical-properties}
\begin{tabular}{ccccc}
\hline
$n_{\rm{0}}$ & $v_{s}$ & $t$ & $M$ & $E_{0}$ \\
 ($10^{-2}$~cm$^{-3}$) & km~s$^{-1}$ & (kyr) & (M$_{\sun}$) & ($10^{51}$ erg) \\
\hline
\hline
1.7--3.1 & 536 -- 608 & 20--25  & 32 -- 62 & 0.37 - 0.75 \\
\hline
\end{tabular}
\end{normalsize}
\end{center}
\end{table}%

\subsubsection{Interior emission}
\label{int-em}
\par We initially fitted the interior emission with a \textit{vapec} model at LMC abundances. However, this resulted in a very poor fit (red. $\chi^{2}>2$) with large residuals at $\sim0.9$ keV due to the Fe L-shell emission lines. Thus, we freed the Fe abundance and re-performed the fit, resulting in a substantially improved fit with $kT=0.76~(0.73-0.79)$~keV, $Z_{\rm{Fe}}>2.01$~$Z/\rm{Z_{\sun}}$, and red. $\chi^{2}=1.08$. The lower limit for the Fe abundance is $\sim4$ times higher than the average LMC value, clearly indicating an overabundance of Fe in the central region, most likely due to Fe-rich ejecta being heated by the reverse shock. Allowing other metal abundances to vary (e.g. O) did not improve the fits and were very poorly constrained. \\

A more realistic model can be applied assuming that the interior gas consists of ejecta only. Hence, we applied thermal plasma models, representative of pure heavy-element plasmas \citep[see][for example]{Kosenko2010}. We performed trial fits with collisional ionisation equilibrium (CIE) models and non-equilibrium ionisation (NEI) models representing the pure metal plasmas and found that, in both cases, a Fe plasma was required to fit the spectrum of the interior emission with a minimal pure O contribution. The CIE models provided acceptable fits to the spectra (reduced $\chi^{2}\approx1$). Similar quality fits were achieved with NEI models with the formal best-fit ionisation timescales determined as $\sim10^{13}$ s cm$^{-3}$, suggestive of a plasma in CIE. However, no constraints could be placed on these values. Given the quality of the trial fits with the CIE models, the interior plasma is suggested to be close to or in CIE. Hence, we model this emission using the \textit{vapec} model in XSPEC. Initial fits showed that the temperature of the O component was not well constrained. Hence, as in \citet{Hughes2003}, we simplified the model by assuming the O and Fe to be co-spatial and constrained their temperatures to be the same. This yielded a best-fit $kT = 0.78^{+0.03}_{-0.03}$~keV. In our spectral fits we found that Fe dominates the interior emission with O having little or no contribution to the spectrum. Indeed, the normalisation of the O component was pegged at zero, demonstrating that it is formally not required. However, we keep the component in the spectral model as the upper limit to the O contribution can allow us to at least determine the maximal O/Fe ratio (see below). We reiterate that the interior plasma is likely in or close to CIE due to the quality of the $\vapec$ model fit  (reduced $\chi^{2}\approx1$, see Table \ref{fit-results}). The $L_{\rm{X}}$ of the interior gas in the 0.2--10~keV range was found to be $\sim2.8\times10^{34}\ \rm{erg\ s}^{-1}$. \\

\par It is possible to identify the likely SN mechanism responsible for \SNR\ by assessing the O/Fe ratio of the interior region. From the SN nucleosynthesis yields of \citet{Iwamoto1999}, the O/Fe ratio is expected to be 0.3--0.7 by number for Type~Ia events and $\sim70$ for core-collapse events. We estimated the maximum allowed O/Fe ratio for \SNR\ using a $\chi^{2}$ contour plot produced for the Fe and O normalisation parameters.  This yielded a maximum ratio of  $<21$ at the 90\% confidence level, slightly higher than the LMC value of 13.2 \citep{Russell1992}. We stress that this upper limit is based on the rather poorly constrained, formally not required O component in the spectral fits. However, it does rule out a CC scenario, indicating \SNR\ likely resulted from a Type~Ia explosion. Another diagnostic for the SN type, is the mass of Fe produced. This can be estimated using Equation 2 of \citet{Kosenko2010}, adapted for Fe. We split this equation into terms which are easily determined from the fit results, given as

\begin{equation}
\label{Fe-calc}
M_{\rm{Fe}}  = (V_{\rm{Fe}} EM_{\rm{Fe}})^{0.5} (n_{\rm{e}}/n_{\rm{Fe}})^{-0.5} m_{\rm{U}} A_{\rm{Fe}}
\end{equation}

\noindent where $V_{\rm{Fe}}$ is the volume occupied by the Fe, $EM_{\rm{Fe}}=n_{e}n_{Fe}V$ is the emission measure of the Fe gas, $n_{\rm{e}}/n_{\rm{Fe}}$ is the electron to Fe-ion ratio, $m_{\rm{U}}$ is the atomic mass unit, and $A_{\rm{Fe}}$ is the atomic mass of Fe. $V_{\rm{Fe}}$ is determined from the dimensions of the central region ($1.4\arcmin\times0.8\arcmin$) and assuming an ellipsoidal morphology with a third semi-axis equal to the mean of the semi-major and semi-minor axes ($1.1\arcmin$). If the actual morphology is oblate or prolate, the volume would be 13\% higher or lower, respectively. $EM_{\rm{Fe}}$ is obtained from the normalisation of the Fe component and is listed in Table \ref{fit-results}. We must estimate the $n_{\rm{e}}/n_{\rm{Fe}}$ ratio. As discussed in \citet{Hughes2003}, there are two plausible scenarios for the number of electrons per Fe ion, depending on the level of admixture of H in the ejecta. If there is no H, then the number of free electrons per Fe ion only depends on the average ionisation state of the each metal contributing electrons. We do not consider the contribution of electrons by the minimal O content of the interior region. For a plasma in CIE at $kT= 0.78$~keV (log $T$= 6.95), the average ionisation is 18.3 for Fe \citep{Shull1982}. Then, from Equation \ref{Fe-calc}, the Fe mass is in the range 1.6--1.8 $M_{\sun}$ for the pure heavy element scenario. The second case proposed by \citet{Hughes2003} is that a similar mass of H is mixed into the metal rich ejecta. Therefore, the number density of Fe over H is 1/56, and the average number of electrons per Fe ion is 74.3. Thus, from Eq. \ref{Fe-calc} we have an Fe mass range of 0.8--0.9 $M_{\sun}$ if there is a comparable mass of H to Fe in the ejecta. The former case gives a value somewhat too high for the explosion of a white dwarf, the latter value being typical of a Type~Ia yield. However, the ejecta might be clumpy, with a filling factor of $\sim0.4$ \citep[see][and references therein]{Kosenko2010}, in which case the Fe content reduces by a factor of $\sqrt{0.4}\approx0.63$ giving 1.0--1.1 $M_{\sun}$ and 0.5--0.6 $M_{\sun}$ for the pure Fe and Fe+H mixture cases, respectively. The large mass of Fe present in the interior further supports a Type~Ia progenitor for \SNR.

\section{Discussion}

The X-ray emission from \SNR\ is notable due to the Fe-rich gas in the interior of the remnant, which is typically indicative of a Type~Ia origin \citep[][and references therein]{Vink2012}. \SNR\ is not the only SNR to exhibit such a feature, with a number of SNRs also characterised by central Fe enhancements in both the LMC: 0454--67.2 \citep{Seward2006}; 0548--70.4 and 0534--69.9 \citep{Hendrick2003}; DEM~L316A \citep{Nish2001,Williams2005}; DEM~L238 and DEM~L249 \citep{Borkowski2006}; and the SMC: DEM~S128, IKT 5, and IKT 25 \citep{Heyden2004,2000A&A...353..129F,2008A&A...485...63F}; SNR~0104--72.3 \citep{Lee2011}. Of these, DEM~L316A, DEM~L238, and DEM~L239 are most similar to \SNR\ as, not only is a faint shell detected surrounding the interior Fe-rich gas, but also the Fe-rich gas is in CIE. However, as argued by \citet{Borkowski2006}, it is likely that many more SNRs fall into this category, only their shells are so faint as to be undetectable by \textit{XMM-Newton}. DEM~S128 \citep{2000A&A...353..129F} and IKT~5 (Roper et al., in preparation) are examples of such SNRs with an Fe-rich gas in CIE but no detected shell. The X-ray structure of such SNRs presents a problem for standard Type~Ia models \citep{Dwarkadas1998,Badenes2003,Badenes2005,Badenes2006}. Assuming that the SN ejecta have expanded in a uniform ambient ISM, and we know from the X-ray analysis that the ambient density is quite low {\bf($\sim10^{-2}\ \rm{cm}^{-3}$)}, then, under the standard Type~Ia models, the central Fe-rich gas should be of low density and the ionisation times are short, which are not observed. \citet{Borkowski2006} demonstrate these scenarios using 1D hydrodynamic simulations using the model of \citet{Dwarkadas1998} and find consistency with the results of  \citet{Badenes2003,Badenes2005,Badenes2006}. \\

\par To explain the X-ray properties of DEM~L238 and DEM~L249, \citet{Borkowski2006} argue that, rather than the SNR expanding into an ideal environment of uniform density, the structure of the ejecta and/or the uniformity of the ambient ISM deviates from the standard Type~Ia models. The best candidates for such a scenario are the `prompt' Type~Ia explosions \citep{Mannucci2006,Sullivan2006,Aubourg2008}. These `prompt' events are produced by relatively massive stars \citep[$\sim3.5-8\ \rm{M}_{\sun}$,][]{Aubourg2008} and are capable of increasing the circumstellar medium (CSM) density via ejection of their stellar envelope prior to the supernova event. Alternatively, the SN ejecta themselves could have stripped the companion of its envelope \citep{Marietta2000}. As stated by \citet{Borkowski2006}, at late stages of SNR evolution, the Fe-contaminated envelope would be evident as dense Fe-rich gas close to the centre of the SNR. Ultimately, this is simply a plausible scenario to explain the observed X-ray morphologies and spectra of \SNR\ and similar remnants. However, it is clear that the standard Type~Ia model fails to explain the properties of these SNRs and some mechanism for a pre-explosion increase in the CSM density is essential. Secondary evidence to support the `prompt' explosion scenario can be found in the stellar content and star-formation history (SFH) of the region, further supporting the scenario described above. \\

To understand the type of stellar environment the progenitor of this SNR came from, we make use of data from the Magellanic Cloud Photometric Survey (MCPS \citealt{2004AJ....128.1606Z}) to construct colour--magnitude diagrams (CMDs) and identify blue stars more massive than $\sim$8~M$_{\sun}$ within a 100~pc ($\sim$6.9\arcmin) radius of \SNR. This allows us to see the prevalence of early-type stars close to the remnant. In the case of \SNR, there are not many early-type stars close to the remnant (16), which is far fewer than around DEM~L205 \citep{2012A&A...546A.109M}, where we found 142 stars and concluded to a CC-SN origin, and more similar to SNRs such as SNR J0530--7007 \citep{2012A&A...540A..25D} with 13 early-type stars or J0527--7104 \citep{2013A&A...549A..99K} with 5 such stars.

\begin{figure}
\centering\includegraphics[trim=10 0 0 0, scale=.53]{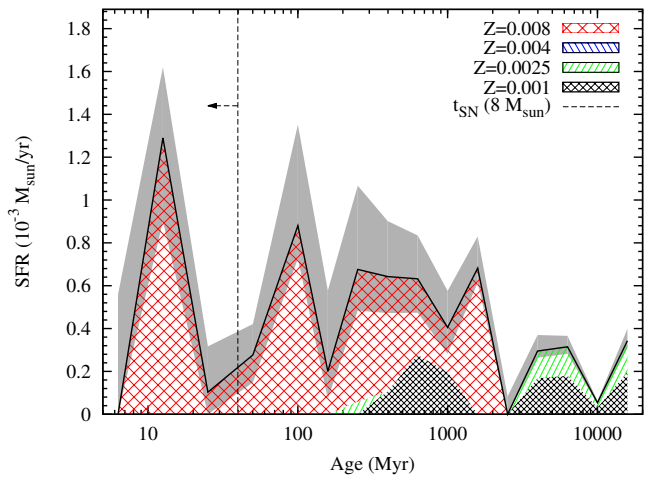}
\caption{Star-formation history around \SNR. We show the star-formation rate (SFR) as a function of look-back time for different metallicity bins, using the map of \citet{2009AJ....138.1243H}. The gray shading denotes the uncertainties on the total SFR. The time scale for a core-collapse SN ($\lesssim40$~Myr) to explode is  indicated by the dashed line.
 \label{sfh_average}}
\end{figure}

Studying the SFH in the neighbourhood of \SNR\ (see Fig. \ref{sfh_average}) as derived from the map of \citet{2009AJ....138.1243H}, we see modest levels of star-formation ($\sim$10$^{-3}$ M$_{\sun}$/yr) peaking 12 Myr ago (accounting for the presence of several early-type stars as described in the previous paragraph), 100 Myr ago, and extending back from 250~Myr to 1~Gyr. The peak at a lookback time of 100~Myr  is consistent with a relatively young progenitor able to produce dense CSM, but we cannot rule out an older progenitor. A `delayed' progenitor older than $\sim$2.4~Gyr (e.g. \citealt{2011MNRAS.412.1508M}) is however less likely.

\citet{Borkowski2006} discussed the possibility of the iron-rich DEM~L238 and DEM~L249 remnants to be those of sub-luminous SN2002cx-like Type~Ia SNe. Recent studies of this peculiar class of SNe \citep{2012ApJ...761L..23J,2013ApJ...767...57F,2013MNRAS.429.2287K} interpret these SNe as failed-detonations of a white dwarf, which leads to relatively low initial kinetic energies ($\sim$0.13$\times$10$^{51}$ erg) and ejecta masses ($\lesssim$ 0.5~M$_{\sun}$). Our estimates of $E_{0}$ and $M_{\rm{Fe}}$ are at odds with these properties, suggesting that \SNR\ is likely not the remnant of a failed-detonation SN. Further studies of centrally iron-rich SNRs and discovery of new members of that class is highly desirable to address the questions of what are the progenitors of these objects, and if and how they differ from the remnants of `standard' Type~Ia SNe.

\section{Summary}

We have undertaken a detailed multi-wavelength study of a new LMC SNR candidate, \SNR, using radio, X-ray and optical data (both imaging and spectroscopy) that confirm we have uncovered a new LMC SNR. With this analysis we infer that \SNR\ is an evolved remnant,
with a large size of $\sim$74$\times$57 pc. We estimate the magnetic field strength to be $\sim$28 $\mu$G and an age of $\sim20-25$ kyr. Based on the determined O/Fe ratio of the interior gas and the mass of Fe ejecta, we conclude that this remnant is of a thermonuclear SN. It is most likely that this remnant was the result of a `prompt' Type~Ia event, sharing similar properties to various other prompt Ia LMC SNRs.\\

\noindent {\bf ACKNOWLEDGEMENTS}\\

\noindent We used the {\sc karma} and {\sc miriad} software package developed by the ATNF. The Australia Telescope Compact Array is part of the Australia Telescope which is funded by the Commonwealth of Australia for operation as a National Facility managed by CSIRO. We thank the Magellanic Clouds Emission Line Survey (MCELS) team for access to the optical images. We thank Dr. David Frew at Macquarie University for useful discussions. This research is supported by the Ministry of Education and Science of the Republic of Serbia through project No. 176005. M.S. acknowledges support by the Deutsche Forschungsgemeinschaft through the Emmy Noether Research Grant SA 2131/1. P. M. and P. K. acknowledge support from the BMWi/DLR grants FKZ 50 OR 1201 and 50 OR 1209, respectively.

\bibliographystyle{mn2e}
\bibliography{real_bib}

\appendix
\label{appA}
\section{X-ray background}
\par In the case of single and double-pixel events, the X-ray background comprises particle-induced and astrophysical components, as well as electronic read-out noise. The particle-induced background of the EPIC consists of three components. The first of these is the quiescent particle background (QPB), a continuum component produced by the direct interaction of penetrating high-energy particles with the detector. Depending on the energy range considered, this can be modelled by a power law not folded through the instrumental response \citep[see][for example]{2004SPIE.5165..112F,Nev2005,Zhang2009}. The second component consists of the instrumental fluorescence lines, which are due to the interaction of high-energy particles with the material surrounding the detector \citep{Lumb2002,Read2003}. Below 5~keV, only the Al K$\alpha$ instrumental fluorescence line at 1.49~keV contributes to the EPIC-pn particle-induced background. Finally, we have residual soft proton (SP) contamination, which is due to SPs travelling through the telescope system to the detector. This component is quite variable with periods of high SP background filtered out of an observation event list during a standard reduction. However, there will always remain some residual SP contamination which escapes the filtering criteria and must be considered. The spectrum of these residual SPs can be described by a power-law not convolved with the instrumental response \citep{Kuntz2008}, and can vary in both magnitude and slope. The electronic read-out noise is very significant at the lowest energies, increasing dramatically below $\sim0.7$ keV. Since we limit our spectral analysis to the 0.4$-$5 keV energy range, we avoid the most affected energy range; however, some increase in the background is expected just above 0.4 keV.

\subsection{Quiescent particle background, instrumental fluorescence lines, and electronic read-out noise}
\par To model the QPB, instrumental fluorescence lines, and the electronic read-out noise we make use of the \textit{XMM-Newton} filter wheel closed (FWC) data, which were collected with the detector shielded from the astrophysical and SP backgrounds. A full discussion on the FWC data of the EPIC-pn is given in \citet{2004SPIE.5165..112F}. The FWC data were screened using the same event filtering criteria as for the observational data, and were also vignetting-corrected using the SAS task \textit{evigweight}. Although the QPB is not subject to vignetting, it is necessary to apply the same correction as to the observational data. For the purposes of spectral fitting, ancillary response files are generated for a flat detector map as the effective area correction has already been applied to the event list.

\begin{figure}
\begin{center}
\resizebox{\hsize}{!}{\includegraphics[trim= 1cm 1.2cm 0.5cm 2.9cm, clip=true, angle=270]{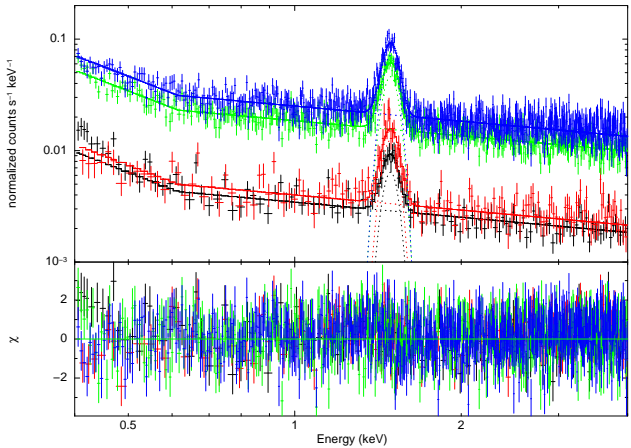}}
\caption{Filter wheel closed spectra extracted from the same source and background regions defined for the X-ray spectral analysis of \SNR. The black, red, green, and blue data points correspond to the SNR centre, centre background, SNR shell, and shell background, respectively.  The best-fit broken power law plus Gaussian line (\textit{bknpower+gauss} in XSPEC) model is indicated by the solid lines ($\Gamma_{1} = 1.91$, $\Gamma_{2} = 0.45$, and $E_{break} = 0.62$~keV). The difference in normalisations between the corresponding source and background spectra is due to the vignetting-correction process (see text).}
\label{qpb-plot}
\end{center}
\end{figure}

To characterise the FWC spectra, and to assess the likely contribution of the QPB to our observational data, we extract spectra from the FWC data from the same regions of the detector as our SNR source and background spectra. However, these spectra must be scaled to account for differences in the integrated QPB count rates of the observational and FWC data. To perform this scaling we considered the 10--13~keV count rates for the entire field-of-view (FOV) of both datasets and determined a scaling factor. Unless a very bright and hard source is present in the observational FOV, the QPB continuum should dominate both datasets at these high energies. Scaling the FWC data using lower energy range count rates is not safe as there is potentially an astrophysical contribution in the observational data. We selected the energy range 10--13~keV for the scaling as \citet{Kat2004} showed a good correlation between the 2--7~keV count rate and the 10--13~keV count rate in the FWC observations. Thus, the 10--13~keV count rates can be used to reliably scale the lower energy QPB continuum. The scaled FWC spectra are shown in Figure \ref{qpb-plot}. As previously mentioned, the background spectra were extracted from regions further off-axis than those of the SNR. The fact that these were obtained from a vignetting-corrected FWC dataset has important effects on the normalisations of the QPB components. The vignetting-correction process operates on every event in the data, regardless of photon or particle origin, according to its position on the detector to account for the variation in effective area of the instrument. However, this correction is photon specific and the particle background is not subject to the same effects. This is clearly evident from Fig. \ref{qpb-plot}, with the background spectra having a larger normalisation than their corresponding source spectra due to the higher vignetting-correction applied to events further off-axis. \\

\par We chose not to subtract these FWC spectra from the observational spectra because, as pointed out by \citet{Kuntz2008} and \citet{Snowden2008}, the fluorescence lines of the EPIC instruments are sufficiently strong that small gain variations and differences in environment between FWC and filter wheel open positions can produce significant differences in the lines and correspondingly large residuals after subtraction. Hence, we chose instead to explicitly model the background in our spectral fits, using the FWC spectra shown in Fig. \ref{qpb-plot} as a basis. We fit models to the FWC spectra simultaneously, allowing only normalisations to vary, and found that the QPB continuum and electronic read-out noise components in the 0.4--4~keV range were best-fit using a broken power law (\textit{bknpower} in XSPEC) not folded through the instrument response, with the energy break ($E_{break}) = 0.62$~keV, the photon index for $E < E_{break}$ ($\Gamma_{1}$) = 1.91, the photon index for $E > E_{break}$ ($\Gamma_{2}$) = 0.45, and reduced $\chi^{2} = 1.02$ for 1363 d.o.f.. Here, $\Gamma_{1}$ represents the electronic read-out noise and $\Gamma_{2}$ the QPB. While at lower energies the electronic read-out noise component requires a more complicated model, in the 0.4-5 keV energy range we consider, a power-law is sufficient. A Gaussian line was also included to account for the Al K$\alpha$ instrumental fluorescence line at 1.49~keV. The fits to the FWC spectra are indicated in Fig. \ref{qpb-plot} as the solid lines. Above 4-5~keV, the defined model no longer applies as the QPB 'flattens'. However, this is unimportant in our analysis as we can easily constrain the QPB model given that this component dominates the emission above 1.5-2~keV. In our spectral analysis of \SNR, the ratios of the normalisations of the source and background QPB components were fixed to the values determined from the QPB spectra. 

\subsection{Soft-proton contamination}
To account for the residual SP contamination in our spectra, we included the appropriate power law model not convolved with the instrument response. However, the fits would consistently return a zero normalisation for the SP component, suggesting that the SP contamination is not significant. This could be expected given the fact that there were no SP flares during the observation. However, we note that a lack of SP flares does not necessarily mean no significant SP contamination. Thus, we ran the diagnostic algorithm of \citet{2004A&A...419..837D}\footnote{See \url{http://xmm2.esac.esa.int/external/xmm_sw_cal/background/epic_scripts.shtml}} which suggested little or no contamination by SPs. A further indication that SP contamination is minimal is that the shell spectrum above 2 keV, where we expect the SP contamination to be most noticeable, can be accounted for by the QPB component, with the normalisation of the QPB components in all our fits agreeing with those of the FWC spectra within their 90\% confidence intervals. For these reasons, we conclude that residual SPs have no significant effect on our spectra and we ignore their contribution in our analysis.

\subsection{Astrophysical background}
\label{A3}
The astrophysical background can usually be modelled with four or fewer components \citep{Snowden2008,Kuntz2010}. These are: the unabsorbed thermal emission from the Local Hot Bubble (LHB, $kT\sim 0.1$~keV), absorbed cool ($kT\sim0.1$~keV) and warm ($kT\sim0.25$~keV) thermal emission from the Galactic halo, and an absorbed power law representing unresolved background cosmological sources \citep[$\Gamma\sim 1.46$,][]{Chen1997}. The cool Galactic halo component likely does not contribute to the astrophysical background in this case as it is absorbed by the foreground Galactic absorbing column and, thus, we do not consider it in our background models. During our analysis, we allowed the LHB and  warm Galactic halo temperatures to vary, with the spectral index of the unresolved background power-law fixed at 1.46. Additionally, it was assumed that the temperature of the thermal components and the surface brightness of the thermal and non-thermal components did not vary significantly between the source and background regions. Thus, the appropriate temperature and normalisation parameters were linked. To model the absorption of the Galactic halo and cosmological background sources we used a photoelectric absorption model in XSPEC. The Galactic halo emission is subject to foreground Galactic absorption ($N_{\rm{H\ Gal}}$), which was fixed at $7\times10^{20}$ cm$^{-2}$ based on the HI maps of \citet{Dickey1990}. The unresolved cosmological background component is not only subject to this Galactic absorption, but also absorption by material in the LMC ($N_{\rm{H\ LMC}}$). This LMC absorption component was incorporated into the fits using a \textit{vphabs} model with abundance at 0.5 solar as appropriate for the LMC \citep{Russell1992}. However, the formal best-fit value of $N_{\rm{H\ LMC}}$ invariably tended to zero and was ultimately fixed as so. The astrophysical background was sufficiently accounted for by our defined model, the best-fit parameters of which were consistent in all of our fits. The temperature of the LHB and warm Galactic halo components were (0.08--0.11)~keV and (0.28--0.33)~keV, respectively, which are compatible the expected values from \citet{Snowden2008} and \citet{Kuntz2010}. 

\label{lastpage}
\end{document}